\documentclass[aps,a4paper,preprint,superscriptaddress,preprintnumbers,floatfix,nofootinbib,amsmath,amssymb]{revtex4}

\usepackage{amstext,amssymb}
\usepackage{amsmath}
\usepackage{graphicx}
\usepackage[hyperfootnotes=true]{hyperref}
\usepackage{color}
\usepackage{comment}
\usepackage{array,multirow}
\usepackage{slashed} 
\usepackage{multirow}
\usepackage{color}
\usepackage{float}
\usepackage{amsfonts}
\usepackage{amsmath}
\usepackage{slashed}
\usepackage{soul}
\usepackage{comment}
\def\beq{\begin{equation}}
\def\eeq{\end{equation}}
\def\bea{\begin{eqnarray}}
\def\eea{\end{eqnarray}}
\newcommand{\bpl}{\beta_{+}}
\newcommand{\bmi}{\beta_{-}}
\newcommand{\mm}{m_{-}}
\newcommand{\mpl}{m_{+}}
\newcommand{\qsq}{q^{2}}
\def\re{{\rm Re}}

\large

\newcommand{\nn}{\nonumber}

\def\s1{\hat s}

\def\U1mt{U(1)_{L_\mu-L_\tau}}

\usepackage{subfigure}
\def\mLb{{m_{\Lambda_b}}}
\def\mmLb{{m^2_{\Lambda_b}}}
\def\mL{{m_{\Lambda}}}
\def\mmL{{m^2_{\Lambda}}}
\def\cl{{\cos\theta_\ell}}

\def\plpl{{+\frac{1}{2}+\frac{1}{2}}}
\def\plmi{{+\frac{1}{2}-\frac{1}{2}}}
\def\mipl{{-\frac{1}{2}+\frac{1}{2}}}
\def\mimi{{-\frac{1}{2}-\frac{1}{2}}}
\def\re{{\rm Re}}  
\def\mC{{\mathcal{C}}}

\def\bl{{\beta_\ell}}

\def\blp{{\beta^\prime_\ell}}

\newcommand{\beas}{\begin{eqnarray*}}
\newcommand{\eeas}{\end{eqnarray*}}

\definecolor{schrift}{RGB}{120,0,0}

\def\ARpe#1{{A^R_{\perp_{#1}}}}  \def\ARpa#1{{A^R_{\|_{#1}}}}

\def\ALpe#1{{A^L_{\perp_{#1}}}}  \def\ALpa#1{{A^L_{\|_{#1}}}}

\def\APpa{{A_{\rm P \|}}} \def\APpe{{A_{\rm P \perp}}}
\def\AsPpa{{A^\ast_{\rm P \|}}} \def\AsPpe{{A^\ast_{\rm P \perp}}}
\def\ASpa{{A_{\rm S \|}}} \def\ASpe{{A_{\rm S \perp}}}
\def\AsSpa{{A^\ast_{\rm S \|}}} \def\AsSpe{{A^\ast_{\rm S \perp}}}
\def\Apat{{A_{\|t}}} \def\Apet{{A_{\perp t}}}
 
\def\AsRpe#1{{A^{R \ast}_{\perp_{#1}}}}  \def\AsRpa#1{{A^{R \ast}_{\|_{#1}}}}

\def\AsLpe#1{{A^{L \ast}_{\perp_{#1}}}}  \def\AsLpa#1{{A^{L \ast}_{\|_{#1}}}}

\usepackage{slashed}

\begin{document}
\title{\bf Unraveling New Physics Effects in  $b \rightarrow s \ell_1 \ell_2$  Transitions with a Model-Independent Perspective}
\author{Aishwarya Bhatta}
\email{aishwaryabhatta@niser.ac.in}
\affiliation{National Institute of Science Education and Research,
An OCC of Homi Bhabha National Institute, Bhubaneswar, Odisha, India}
\author{Dhiren Panda}
\email{pandadhiren530@gmail.com}
\author{Rukmani Mohanta}
\email{rmsp@uohyd.ac.in}
\affiliation{School of Physics,  University of Hyderabad, Hyderabad-500046,  India}
\begin{abstract}
Motivated by recent anomalies in observables associated with flavor-changing neutral current (FCNC) transitions, specifically $b \rightarrow s \ell^+ \ell^-$ processes, we present a comprehensive analysis of lepton flavor-violating (LFV) decay modes mediated by $b \rightarrow s \ell_1 \ell_2$ transitions with $\ell_1 \neq \ell_2$. While such LFV processes are forbidden within the Standard Model (SM), they naturally arise in several of its extensions, including models featuring additional vector-like fermions and extra $Z'$ bosons. Employing the most general effective Hamiltonian for $b \rightarrow s \ell_1 \ell_2$ transitions, we derive the angular distributions of the relevant decay modes. Adopting a model-independent framework, we systematically study the LFV decays $B \rightarrow K^* \ell_1 \ell_2$, $B_s \rightarrow \phi \ell_1 \ell_2$, $B \rightarrow K_2^* \ell_1 \ell_2$, and $\Lambda_b \rightarrow \Lambda \ell_1 \ell_2$. Although LFV mesonic decays have been widely explored, the corresponding baryonic decays remain comparatively under-investigated. We provide bounds on branching ratio ($\mathcal{B}$), forward-backward asymmetry ($\mathcal{A}_{FB}$), and longitudinal lepton polarization fraction ($\mathcal{F}_L$). Furthermore, considering the projected sensitivities of the LHCb upgrade and Belle II experiments, we estimate upper limits for these observables, offering promising avenues for probing new physics in these LFV channels.

\end{abstract}
\maketitle
\flushbottom
\section{Introduction} 
The SM has proven remarkably effective in describing a broad range of physical phenomena, particularly at the electroweak scale. Nevertheless, it is now widely acknowledged that the SM is incomplete, as it fails to explain several critical aspects of nature. These include the origin of neutrino masses and mixing, the composition of dark matter, and the matter-antimatter asymmetry observed in the universe. To probe these open questions, flavor physics serves as a key tool. In particular, the analysis of rare decays mediated by FCNC transitions provides a promising avenue for discovering NP beyond the SM. Significant discrepancies between experimental observations and SM predictions in these processes would signal potential contributions from NP.

Among these rare processes, the FCNC transitions $b \to s \ell \ell$ are of particular interest due to their rich phenomenology and the availability of theoretically clean observables. Ratios pertaining to lepton flavor universality (LFU), such as $R_K$ and $R_{K^*}$—defined as the ratios of branching fractions $\mathcal{B}(B \to K^{(*)} \mu \mu)/\mathcal{B}(B \to K^{(*)} e e)$—are especially useful, as hadronic uncertainties cancel to a large extent. Similarly, LFU observables in charged-current processes like $R_D$ and $R_{D^*}$ (in $b \to c \ell \nu$ transitions) show a deviation of around $3\sigma$ from SM predictions. Years of collaboration between theoretical and experimental physicists have significantly refined our ability to test the SM.

Earlier measurements by the LHCb collaboration reported a $3.1\sigma$ deviation in $R_K$ within the $q^2$ range $[1.1,6.0]~{\rm GeV}^2$~\cite{LHCb:2021trn}. Similarly, $R_{K^*}$ measurements in the low-$q^2$ bins $[0.045,1.1]$ and $[1.1,6.0]~{\rm GeV}^2$ showed deviations of $2.2\sigma$ and $2.5\sigma$, respectively \cite{LHCb:2017avl,LHCb:2020lmf}. However, the situation changed notably with the updated LHCb results announced in December 2022 \cite{LHCb:2022qnv,LHCb:2022zom}, which reported:
\begin{align}
R_K &= 
\begin{cases}
0.994^{+0.090}_{-0.082}~(\text{stat})^{+0.027}_{-0.029}~(\text{syst}), & 0.045 \le q^2 \le 1.1~{\rm GeV}^2, \\
0.949^{+0.042}_{-0.041}~(\text{stat})^{+0.023}_{-0.023}~(\text{syst}), & 1.1 \le q^2 \le 6.0~{\rm GeV}^2,
\end{cases} \\
R_{K^*} &= 
\begin{cases}
0.927^{+0.093}_{-0.087}~(\text{stat})^{+0.034}_{-0.033}~(\text{syst}), & 0.045 \le q^2 \le 1.1~{\rm GeV}^2, \\
1.027^{+0.072}_{-0.068}~(\text{stat})^{+0.027}_{-0.027}~(\text{syst}), & 1.1 \le q^2 \le 6.0~{\rm GeV}^2.
\end{cases}
\end{align}
These updated values are more or less consistent with the SM predictions, with deviations reduced to approximately $0.2\sigma$. Despite this, the presence of NP cannot be definitively excluded, as anomalies persist in other observables. For example, measurements of the angular observable $P_5^{\prime}$ from LHCb~\cite{LHCb:2013ghj, LHCb:2015svh} and ATLAS~\cite{ATLAS:2018gqc} report deviations up to $3.3\sigma$, while CMS~\cite{CMS} and Belle~\cite{Belle:2016xuo} show discrepancies of about $1\sigma$ and $2.6\sigma$, respectively.

These inconsistencies collectively known as $B$-anomalies, which highlight the potential involvement of NP, especially in semileptonic $B$ meson decays~\cite{Aaij:2021vac, Aaij:2020ruw, Aaij:2020nrf, Aaij:2014ora, Aaij:2017vbb, Aaij:2019wad, Aaij:2015oid, Lees:2012xj, Lees:2013uzd, Aaij:2015yra, Hirose:2016wfn, Hirose:2017dxl, Aaij:2017uff, Aaij:2017deq, Abdesselam:2019dgh}. Unlike lepton-flavor-conserving decays, LFV transitions offer a particularly clean probe for NP, as they are forbidden in the SM. 

Various LFV processes have been explored in both the charged lepton sector (e.g., $\ell_i \to \ell_j \gamma$, $\ell_i \to \ell_j \ell_k \bar{\ell}_k$) and in $B$ meson decays via $b \to s \ell_i \ell_j$ transitions~\cite{Lee:2015qra, Altmannshofer:2015mqa, Crivellin:2015mga, Alonso:2015sja}. Experimentally, only upper bounds exist so far. LHCb has set an upper limit $\mathcal{B}(B_s \to e^\pm \mu^\mp) < 6.3 \times 10^{-9}$ at $90\%$ C.L., and $\mathcal{B}(B \to \tau \mu) < 4.2 \times 10^{-5}$~\cite{LHCb:2019ujz}. LFV searches in semileptonic modes like $B \to K \ell_1 \ell_2$ yielded $\mathcal{B}(B \to K \mu^- e^+) < 7 \times 10^{-9}$~\cite{LHCb:2019bix}, and for vector meson modes $B \to K^* \mu e$ and $B \to \phi \mu e$, limits are $6.8 \times 10^{-9}$ and $10.1 \times 10^{-9}$, respectively. Despite the experimental complexity introduced by $\tau$ leptons in the final state, Belle has reported $\mathcal{B}(B^+ \to K^+ \mu^\pm \tau^\mp) < 3.9 \times 10^{-5}$~\cite{Belle:2022pcr}. LHCb searches for $B \to K^* \mu^\pm \tau^\mp$ yielded $\mathcal{B}(B \to K^* \tau^+ \mu^-) < 1.0 \times 10^{-5}$ and $\mathcal{B}(B \to K^* \tau^- \mu^+) < 8.2 \times 10^{-6}$~\cite{LHCb:2022wrs}.

$B$ anomalies fall into two major categories: (i) deviations in $\tau$ vs. light-lepton universality in $b \to c \ell \nu$ transitions, and (ii) deviations in FCNC-mediated $b \to s \ell^+ \ell^-$ processes. These suggest the existence of NP particles and have prompted extensive theoretical and experimental exploration. Proposed NP models range from heavy vector and scalar mediators to leptoquarks and non-universal $Z'$ bosons~\cite{Calibbi:2017qbu, Barbieri:2017tuq, Blanke:2018sro, DiLuzio:2018zxy, Faber:2018qon, Heeck:2018ntp, Angelescu:2018tyl, Schmaltz:2018nls, Greljo:2018tzh, Fornal:2018dqn, Baker:2019sli, Cornella:2019hct, DaRold:2019fiw, Bordone:2017bld, Bordone:2018nbg, Bordone:2019uzc, Marzocca:2018wcf, Becirevic:2018afm, Bigaran:2019bqv, Crivellin:2019dwb, Saad:2020ihm, Gherardi:2020qhc, Babu:2020hun, Crivellin:2017zlb, Buttazzo:2017ixm, Bordone:2020lnb, Cornella:2021sby, Marzocca:2021azj, Greljo:2021xmg, Davighi:2021oel, Alvarado:2021nxy}. Common to these models is the prediction that violations of LFU may also imply LFV in $B$ and $\tau$ decays. Currently, experimental upper bounds on LFV modes place stringent constraints on these NP scenarios. Future data will be crucial in confirming or refuting these possibilities.

To investigate LFV in $b \to s \ell_1 \ell_2$ transitions, both mesonic and baryonic decays must be considered. Despite sharing the same underlying partonic process, they offer complementary perspectives on NP. For instance, while short-distance NP effects are absent in $\Lambda_b \to \Lambda \mu^+ \mu^-$ decays~\cite{Blake:2019guk, Bediaga:2018lhg}, such effects appear to influence angular observables in $B \to K^* \mu^+ \mu^-$ decays~\cite{Alguero:2021anc, Altmannshofer:2021qrr, Hurth:2021nsi, Ciuchini:2019usw}. This suggests that decays like $\Lambda_b \to \Lambda \ell_1 \ell_2$ can provide additional insights alongside mesonic channels such as $B^+ \to K^+ \ell_1 \ell_2$ and $\bar{B}_s \to \ell_1 \ell_2$. Moreover, the spin structure of the baryonic decays leads to a richer set of hadronic matrix elements.

Although so far there is no direct evidence of LFV, many theoretical models involving the influence of FCNC mediated by the $Z$ boson \cite{Mohanta:2010yj, Nayek:2020lpg}, non-universal $Z'$ models \cite{Crivellin:2015era, Farzan:2015hkd}, leptoquark models \cite{Das:2019omf},  Minimal Supersymmetric Standard Model (MSSM) \cite{Crivellin:2018mqz, Dedes:2008iw}, and other NP models \cite{Becirevic:2016zri} have been proposed to explain them. Model-independent analyses for only tau decays have been conducted in Ref.~\cite{ Dassinger:2007ru}. This work focuses on LFV decays mediated by $b \to s \ell_1 \ell_2$ transitions and evaluates $\mathcal{B}$, $\mathcal{A}_{FB}$, and $\mathcal{F}_L$ in the channels $B \to K^* \ell_1 \ell_2$, $B \to \phi \ell_1 \ell_2$, $B \to K_2^* \ell_1 \ell_2$, and $\Lambda_b \to \Lambda \ell_1 \ell_2$. Constraints on NP couplings are derived from current experimental bounds ${\cal B}(B^+ \to K^+ \tau^\mp \mu^\pm)$ and ${\cal B}(B_s \to\tau^\mp \mu^\pm)$.
The remainder of this paper is structured as follows. Section~\ref{subsec:effHam} introduces the effective Hamiltonian and theoretical framework for $b \to s \ell_1 \ell_2$ transitions. Section~\ref{sec:kstar-phi} presents the numerical analysis for $B \to K^* \ell_1 \ell_2$ and $B \to \phi \ell_1 \ell_2$ decays. Section~\ref{sec:k2star} focuses on $B \to K_2^* \ell_1 \ell_2$, and Section~\ref{sec:lambda} discusses results for $\Lambda_b \to \Lambda \ell_1 \ell_2$. A summary of key findings is provided in Section~\ref{sec:summary}.
\section{Effective Hamiltonian and theoretical aspects \label{subsec:effHam}}

We commence by employing the  effective Hamiltonian to describe the LFV in $b \to s\ell_1^+\ell_2^-$ transition:
\begin{equation}\label{eq:Heff1}
\mathcal{H}^{\rm eff} = - \frac{G_F \alpha_{\rm em} }{\sqrt{2} \pi} V_{ts}^* V_{tb} \sum_i\bigg( \mC_i \mathcal{O}_i +  \mC^\prime_i \mathcal{O}^\prime_i \bigg)\, ,
\end{equation}
where $G_F$ is the Fermi constant, $\alpha_{\rm em}$ is the fine-structure constant, $V_{ts}$ and $V_{tb}$ are the Cabibbo-Kobayashi-Maskawa (CKM) matrix elements, and $i = S, P, V, A$ correspond to scalar, pseudo-scalar, vector and axial-vector operators, which read as
\begin{eqnarray}\label{eq:opbasis}
\begin{split}
&\mathcal{O}_S^{(\prime)} = \big[\bar{s}P_{R(L)}b \big]\big[\bar{\ell_2}\ell_1 \big]\, ,\quad \mathcal{O}_P^{(\prime)} = \big[\bar{s}P_{R(L)}b \big]\big[\bar{\ell_2}\gamma_5\ell_1 \big]\, \\
&\mathcal{O}^{(\prime)}_V = \big[\bar{s}\gamma^\mu P_{L(R)}b \big]\big[\bar{\ell_2}\gamma_\mu\ell_1 \big]\, 
 ,\quad \mathcal{O}^{(\prime)}_{A} = \big[\bar{s}\gamma^\mu P_{L(R)}b \big]\big[\bar{\ell_2}\gamma_\mu\gamma_5\ell_1 \big]\, ,
\end{split}
\end{eqnarray}
 where $P_{L,R}=(1\mp\gamma_5)/2$ are the chiral projectors. It is important to highlight that the electromagnetic penguin operator $\mathcal{O}_7$, which is involved in $b \to s\ell\ell$ decays, is characterized by,
\begin{equation}
   { \mathcal{O}_{7} = \frac{e}{16 \pi^2}m_b(\bar{s}\sigma_{\mu\nu}P_R b)F^{\mu\nu} \;.}
\end{equation}
However, it is crucial to note that this operator cannot induce lepton flavor-violating contributions because of the universal nature of electromagnetic interactions. The coefficients $\mC_{S, P, V, A}^{(\prime)}$ represent short-distance Wilson coefficients that are absent in the SM but can take non-zero values in various scenarios outside the SM. In the SM, $\ell_1$ and $\ell_2$ are typically leptons of the same flavor, often denoted as $\ell$, and the corresponding operators $\mathcal{O}_{V, A}$ are conventionally labeled as $\mathcal{O}_{9,10}$ with their respective Wilson coefficients denoted as $\mathcal{C}_{9,10}$.
\subsection{Constraints on New Couplings}
First, we consider the constraints on several combinations of Wilson coefficients from measurements of mesonic LFV decays. The quark level transition for the mesonic $\bar{B}_s \to\ell_1^-\ell_2^+$ and $B\to K \ell_1^-\ell_2^+$ decays is the same as the processes we are looking for. So we can use the upper limit on the combinations of  Wilson coefficients extracted from these decay modes and can obtain  the upper limits on various observables of  $B_{s} \to \{ K^*, \phi\} \ell_1 \ell_2$, $ B \to K_{2}^{*}\ell_1 \ell_2$ and $\Lambda_{b} \to \Lambda \ell_1 \ell_2 $ 
 processes. We consider the branching ratios of the decay modes $\bar{B}_s \to\ell_1^-\ell_2^+$ and $B\to K \ell_1^-\ell_2^+$, for which the experimental upper limits at $90\%$ C.L. are reported in Table~\ref{tab:UpperBounds-inputs}. 
\begin{table}[htb]
\begin{center}
\begin{tabular}{||c |c||}
\toprule
Observable & Upper Bound \\ \hline \hline
$\mathcal{B}(B^+\to K^+ \tau^{+} \mu^{-})$ & ~~~~~~ $3.9 \times 10^{-5}$\, \cite{Aaij:2020mqb} \\   \hline
$\mathcal{B}(B^+ \to K^+  \tau^{-}\mu^{+})$~~ & ~~~~~ $4.5 \times 10^{-5}$\, \cite{Lees:2012zz} \\ 
\hline
$\mathcal{B}(\bar{B}_s \to \tau^{\mp} \mu^{\pm} )$ & ~~~~~ $ 3.5
\times 10^{-5}$\, \cite{Aaij:2019okb} \\
 \hline \hline
\end{tabular}
\caption{Experimental upper limits for LFV $B$ decays at $90\%$ C.L..} 
\label{tab:UpperBounds-inputs}
\end{center}
\end{table}
Utilizing Eqn.~(\ref{eq:Heff1}), the branching ratios for the corresponding decay channels can be written as follows:
 \begin{equation}\label{eq:BrB2Kll}
\begin{aligned}
\mathcal{B}(B^+\to K^+\ell_1^-\ell_2^+) &= 10^{-8} \bigg\{c^P_{\ell_1\ell_2}\left|C_{P_+}\right|^2+ c^{10+}_{\ell_1\ell_2}\left|C_{10_+} \right|^2 + c^S_{\ell_1\ell_2}\left|C_{S_+}\right|^2  \\[2pt]
 &  +c^{9+}_{\ell_1\ell_2}\left|C_{9_+} \right|^2  + c^{P10}_{\ell_1\ell_2}\,\mathrm{Re}[C_{P_+}^{*} C_{10_+}+c^{S9}_{\ell_1\ell_2}\,\mathrm{Re}[C_{S_+}^{*}  C_{9_+}]
 \bigg\}\,, 
\end{aligned}
\end{equation}
and
\begin{equation}\label{eq:BrBs2ll}
\begin{aligned}
\mathcal{B}(\bar{B}_s\to\ell^-_1 \ell^+_2)=&\,\frac{\alpha_\text{em}^2G_F^2 |V_{tb}V_{ts}^*|^2}{m_{B_s}^3} \frac{\tau_{B_s}}{64\pi^3} f_{B_s}^2 \, \lambda^{1/2}(m_{B_s}^2,m_{\ell_1}^2,m_{\ell_2}^2)\\
&\left\{[m_{B_s}^2-(m_{\ell_1}-m_{\ell_2})^2]\left|\frac{m_{B_s}^2}{m_b+m_s}{C}_{P_-}+(m_{\ell_1}+m_{\ell_2}){C}_{10_-}\right|^2\right. \\
&+\left.[m_{B_s}^2-(m_{\ell_1}+m_{\ell_2})^2]\left|\frac{m_{B_s}^2}{m_b+m_s}{C}_{S_-}+(m_{\ell_1}-m_{\ell_2}){C}_{9_-}\right|^2\right\} \,.
\end{aligned}
\end{equation}
Both Eqns.~(\ref{eq:BrB2Kll}) and ~(\ref{eq:BrBs2ll}) agree with previous results in the literature \cite{Becirevic:2016oho,Gratrex:2015hna}.
\begin{table}[htb]
\begin{center}
\begin{tabular}{||c  c c c c c||}
\hline\hline
$c_{\ell_1\ell_2}^{9+}$ & $c_{\ell_1\ell_2}^{10+}$ & $c_{\ell_1\ell_2}^{S}$ & $c_{\ell_1\ell_2}^{P}$ & $c_{\ell_1\ell_2}^{S9}$ & $c_{\ell_1\ell_2}^{P10}$ \\
\hline
 $\phantom{-}1.09$ &  $\phantom{-}1.14$ & $\phantom{-}1.47$ &  $\phantom{-}1.58$ & ~ $-1.35$ &  $\phantom{-}1.66$ \\
 \hline\hline
\end{tabular}
\caption{The predictions for the coefficients governing $B^+\to K^+ \ell_1^-\ell_2^+$ decays $(\ell_1^- = \mu^-,\,\ell_2^+ = \tau^+)$ based on the hardonic form factors from Ref.~\cite{Bouchard:2013eph,Gubernari:2018wyi}.}
\label{tab:BtoKtaumu}
\end{center}
\end{table}
Here we have adopted the notation $C_{Y\pm}=C_{Y}\pm C_{Y}^{\prime}$. By using the mass values from the Particle Data Group (PDG) \cite{Zyla:2020zbs}, the CKM factors obtained from the UT-fit collaboration \cite{UTFit}, the decay constant $f_{B_s}=215\mathrm{MeV}$ as provided in \cite{Balasubramanian:2019net}, and relying on Lattice QCD/Light Cone Sum Rule results from Refs.~\cite{Bouchard:2013eph} and \cite{Gubernari:2018wyi}, we have determined the coefficients $c_{\ell_1\ell_2}^i$ in Eqn~(\ref{eq:BrB2Kll}), as displayed in Table~[\ref{tab:BtoKtaumu}]. It's worth mentioning that these coefficients are not influenced by the charges of the leptons, with the exception of $c_{\ell_1\ell_2}^{S9}$. It is important to note that the coefficient $c_{\ell_1\ell_2}^{S9}$ is directly proportional to the difference in masses between $\ell_1$ and $\ell_2$. As a result, this coefficient changes its sign based on the charge of the heavier lepton in the process. We should emphasize that the values presented in Table~[\ref{tab:BtoKtaumu}] are significantly influenced by the choice of the electromagnetic coupling constant $\alpha_\mathrm{em}$. In this analysis, we have used $\alpha_\mathrm{em}=1/133$. However, it's worth noting that a different value for $\alpha_\mathrm{em}$ can be applied by appropriately scaling the $c_{\ell_1\ell_2}^i$ coefficients.\\
 To determine the NP parameter space consistent with current experimental data, we perform a $\chi^2$ analysis. The $\chi^2$ function is defined as
\begin{equation}
\chi^2(C_{\text{NP}}) = \sum_i \frac{\left(\mathcal{O}_i^{\text{Th}}(C_{\text{NP}}) - \mathcal{O}_i^{\text{Exp}}\right)^2}{(\Delta \mathcal{O}_i^{\text{Exp}})^2 + (\Delta \mathcal{O}_i^{\text{SM}})^2},
\label{eq:chi2}
\end{equation}
where $\mathcal{O}_i^{\text{Th}}$ and $\mathcal{O}_i^{\text{Exp}}$ denote the theoretical prediction and the experimental central value of the $i$-th observable, respectively. The denominator accounts for both experimental and SM theoretical uncertainties. We employ the experimental upper limits provided in Table [\ref{tab:UpperBounds-inputs}] and Eqns. (\ref{eq:BrB2Kll}) and (\ref{eq:BrBs2ll}) to constrain various combinations of NP couplings . It is important to note that for the decay modes $B^+ \to K^+ \mu^\pm \tau^\mp$ and $B_s \to \mu^\pm \tau^\mp$, only 90\% confidence level (C.L.) upper bounds on the branching ratios are available. To incorporate these into the $\chi^2$ analysis, we model the corresponding observable as having a central value of zero and an uncertainty given by $\text{U.L.}/1.645$, where U.L. denotes the upper limit. Since these processes are highly suppressed in the Standard Model, the associated theoretical uncertainty from the SM is taken to be negligible. The NP parameter space is then obtained by minimizing the $\chi^2$ function. The analysis is carried out using the best-fit values of the Wilson coefficients, given by $C_9 = -55.361$, $C_{10} = 2.3\times 10^{-2}$, $C_S = -8.941$, and $C_P =6.763$.

It's important to reiterate that we do not include $\tau e$ decays in this analysis, as the constraints derived from these decays resemble those from the $\tau \mu$ channel. Additionally, for the sake of simplicity, we do not account for the $\mathcal{O}^{\prime\ell_1\ell_2}_i$ operators. This choice is made due to the fact that these operators are less favorable when attempting to fit data from $b\to s \ell \ell$ transitions \cite{Alguero:2021anc,Altmannshofer:2021qrr,Hurth:2021nsi,Ciuchini:2019usw,Lancierini:2021sdf}. Nevertheless, it's worth emphasizing that baryonic channels exhibit different dependencies on the primed operators compared to mesonic channels. This distinction could become relevant as scenarios involving these operators gain prominence in addressing the $B$ anomalies.\\
We explore two distinct 2-dimensional scenarios, each of which allows only specific combinations of NP Wilson coefficients to have non-zero values. These scenarios encompass, like the first scenario, $C_9^{\ell_1\ell_2}$ and $C_{10}^{\ell_1\ell_2}$, with the constraint that only these two coefficients are allowed to be non-zero.  The second scenario is inspired by the Standard Model Effective Field Theory (SMEFT), where we set $C_9^{\ell_1\ell_2}=-C_{10}^{\ell_1\ell_2}$ and $C_S^{\ell_1\ell_2}=-C_P^{\ell_1\ell_2}$ as the specific constraints on the coefficients.\\
  $ \textbf{Scenario ~ I }$ : ~ $C_9^{\ell_1 \ell_2}\neq 0, C_{10}^{\ell_1 \ell_2} \neq 0, ~C_S^{\ell_1 \ell_2} =C_P^{\ell_1 \ell_2} =0  $ \\
$ \textbf{Scenario ~II}$ :~ $C_9^{\ell_1 \ell_2}=-C_{10}^{\ell_1 \ell_2},~ C_{S}^{\ell_1 \ell_2}=-C_P^{\ell_1 \ell_2}$\\

\begin{figure}[h!]
\begin{tabular}{cccc}
	\centering
	\includegraphics[width=0.25\textwidth]{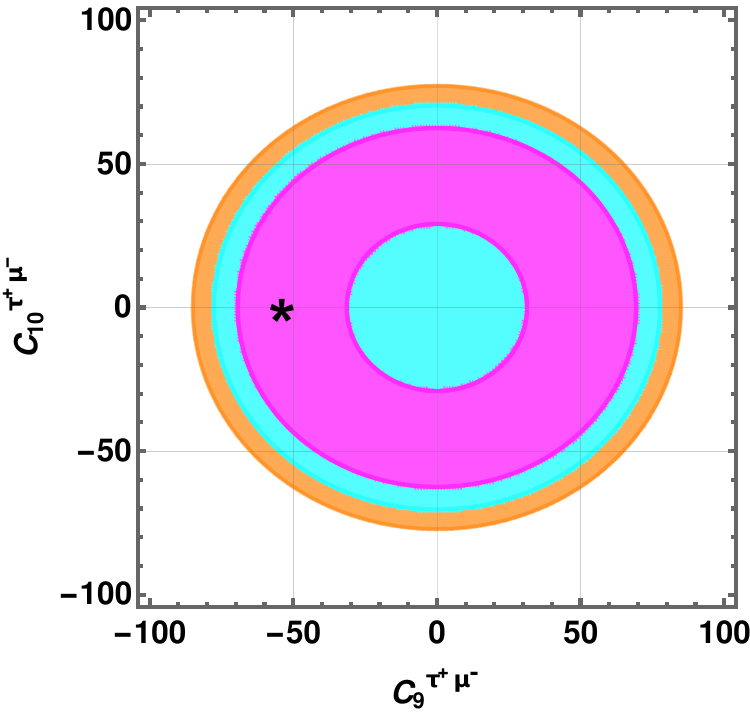} &
  	\includegraphics[width=0.25\textwidth]{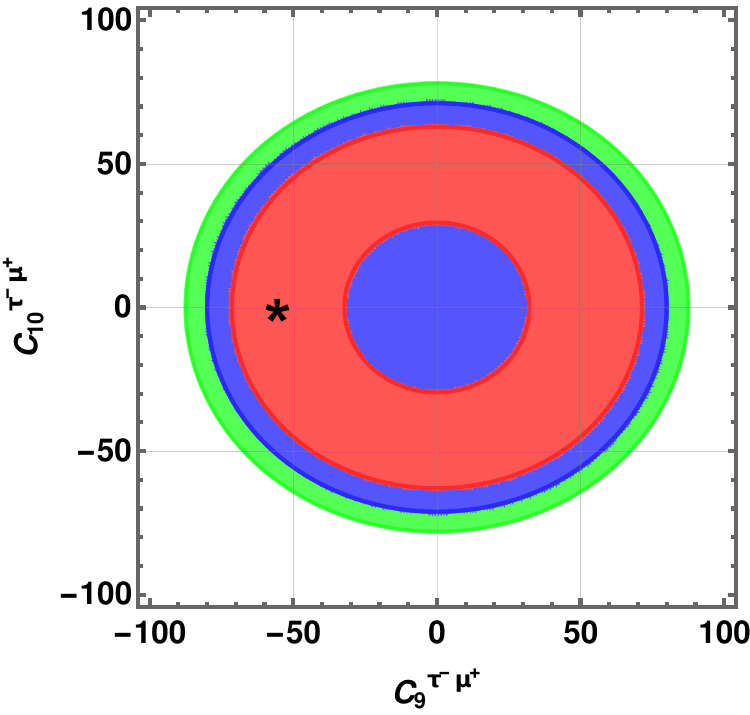} &
  	 \includegraphics[width=0.25\textwidth]{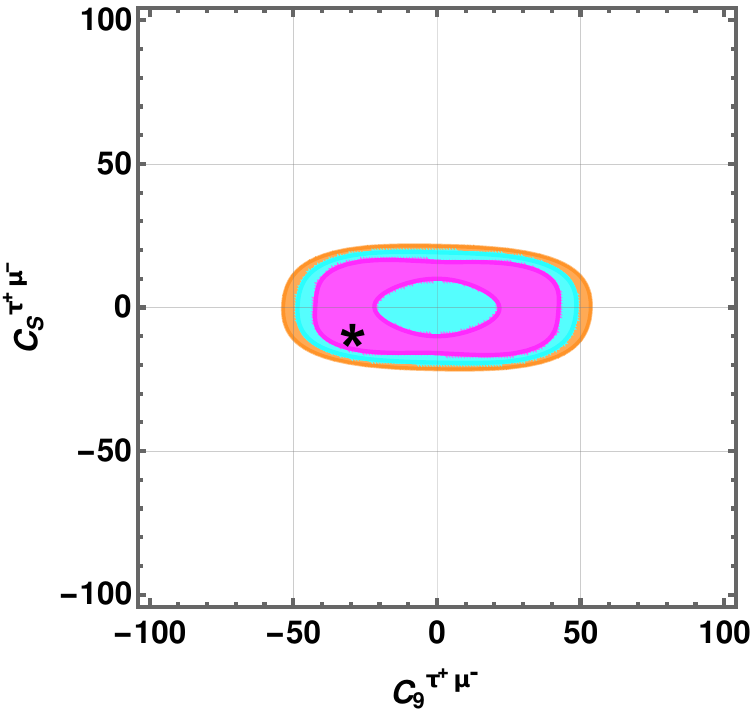} & 
  	\includegraphics[width=0.25\textwidth]{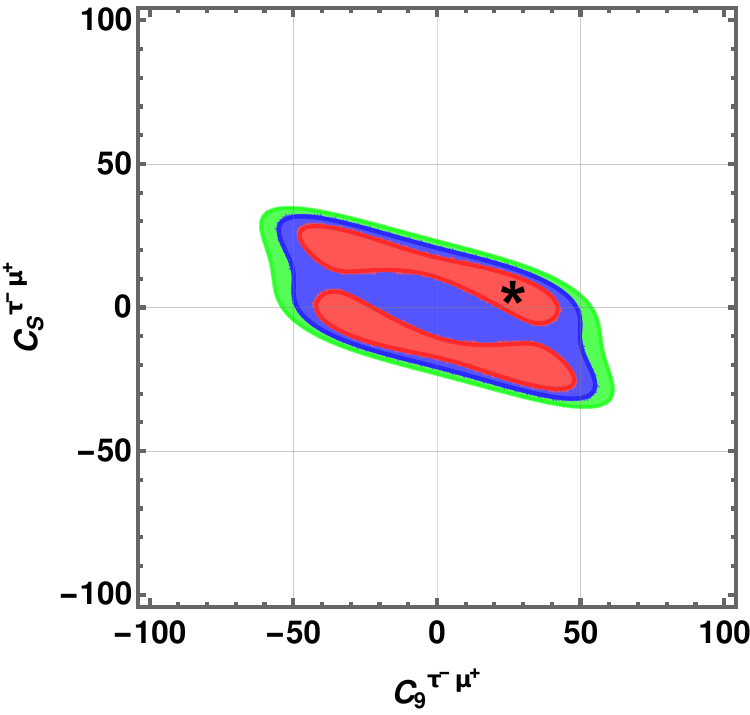} \\
  	\end{tabular}
    \caption{ The allowed parameter space is determined by combining the constraints on various Wilson coefficients derived from the $90\%$ confidence level upper limits of the $\bar{B}_s \to \ell_1^- \ell_2^+$ and $B \to K \ell_1^- \ell_2^+$ decays. In the $\tau^+ \mu^-$ channel, the magenta, cyan, and orange regions represent the 1$\sigma$, 2$\sigma$, and 3$\sigma$ error bands, respectively, while for the $\tau^- \mu^+$ mode, the corresponding bands are shown in red, blue, and green.
}
	\label{fig::ModelIndep-constrainstau}
\end{figure}

The obtained bounds for $\tau^+\mu^-$ and $\mu^+ \tau^-$ finals states are given in Fig.~[\ref{fig::ModelIndep-constrainstau}].The top two plots of the figure show  $C_9$ versus $C_{10}$ and $C_S$ versus $C_P$, while setting all others to zero, which are independent of the charge configuration in the final state, we only consider the strongest bound of $\bar{B}_s \to \tau^+\mu^-$ and $B\to K \tau^+\mu^-$ decay processes. Since the interference between $C_9^{\ell_1\ell_2}$ and $C_S^{\ell_1\ell_2}$ relies on the charge configuration of the leptons in the final state, we have generated plots in the bottom panel for both the $\tau^+\mu^-$ and $\tau^-\mu^+$ final state. Upon comparing the plots in Figure~[\ref{fig::ModelIndep-constrainstau}], we observe substantial disparities between the $\tau^+\mu^-$ and $\tau^-\mu^+$. Consequently, we emphasize the importance of conducting separate analyses for these distinct final states.
\section{Decay distribution of $ B\to K^*\ell_1\ell_2 $ and $B\to \phi \ell_1\ell_2$ $\label{sec:kstar-phi}$}

These processes occur through two different decay channels: $B\to K^*(\to K\pi) \ell_1 \ell_2$ and $B_s\to \phi(\to K\bar K) \ell_1 \ell_2$. Since the angular distribution expression for the latter decay can be derived by straightforward substitutions of the corresponding mass and form factors in the expression of the former decay mode, our primary focus will be on $\bar B\to \bar K^*(\to K^-\pi^+) \ell_1^-\ell_2^+$. Here we adopt the kinematics conventions which are fine-tuned to align with the standards adhered to in experiments conducted at the LHC \cite{Aaij:2015oid}. Additional details regarding the kinematics of this process are provided in the Appendix of this paper. In addition to $\theta_\ell$, we also consider the parameter $\theta_K$, which signifies the angle between the decay axis oriented along the $-z$ direction and the trajectory of the $K^-$ particle in the rest frame of $\bar K^*$. This is illustrated in Figure~[ \ref{fig:angles}]. Furthermore, we use $\phi$ to denote the angle between the planes defined by the $K\pi$ and $\ell_1^-\ell_2^+$ systems.\\
In this scenario, the hadronic matrix elements involve a more extensive set of form factors, including:
\begin{align}\label{def:FFV}
\langle \bar{K}^\ast(k)|\bar{s}\gamma^\mu(1-\gamma_5) b|\bar{B}(p)\rangle &= \varepsilon_{\mu\nu\rho\sigma}\varepsilon^{\ast\nu}p^\rho k^\sigma \frac{2 V(q^2)}{m_B+m_{K^\ast}}-i\Big{[} \varepsilon_\mu^\ast(m_B+m_{K^\ast})A_1(q^2)\nonumber\\[.3em] 
&-i(p+k)_\mu (\varepsilon^\ast \cdot q)\frac{A_2(q^2)}{m_B+m_{K^\ast}}- q_\mu(\varepsilon^\ast \cdot q) \frac{2 m_{K^\ast}}{q^2}[A_3(q^2)-A_0(q^2)]\Big{]}, \nonumber\\[.7em] 
\langle \bar{K}^\ast(k)|\bar{s}\sigma_{\mu\nu} q^\nu(1-\gamma_5) b|\bar{B}(p)\rangle &= 2 i \varepsilon_{\mu\nu\rho\sigma} \varepsilon^{\ast\nu}p^\rho k^\sigma T_1(q^2)-[(\varepsilon^\ast \cdot q)(2p-q)_\mu - \varepsilon_\mu^\ast(m_B^2-m_{K^\ast}^2)]T_2(q^2)\nonumber\\[.3em] 
&-(\varepsilon^\ast \cdot q)\Big{[} \frac{q^2}{m_B^2-m_{K^\ast}^2}(p+k)_\mu - q_\mu \Big{]}T_3(q^2),
\end{align}
where $\varepsilon_\mu$ is the polarization vector of $K^\ast$, and $A_{1,2,3}(q^2)$ are the form factors having the relation $2 m_{K^\ast} A_3(q^2)=(m_B+m_{K^\ast})A_1(q^2)-(m_B-m_{K^\ast})A_2(q^2)$.
The full angular distribution of the above decay processes is~\footnote{Please notice that the convention used in eq.~(\ref{def:FFV}) is such that $\varepsilon_{0123}=+1$. }
\begin{equation}
\dfrac{\mathrm{d}^4 {\cal B} ({B}\to\bar{K}^{\ast}\to (K\pi) \ell_1^-\ell_2^+)}{\mathrm{d}q^2\mathrm{d}\cos \theta_\ell \mathrm{d}\cos \theta_K \mathrm{d}\phi} = \dfrac{9}{32\pi}I(q^2,\theta_\ell,\theta_K,\phi),
\end{equation}
with
\begin{align}
I(q^2,\theta_\ell,\theta_K,\phi) = & I_1^c(q^2)+[I_1^s(q^2)-I_1^c(q^2)]\sin^2\theta_K +[I_2^c(q^2)+[I_2^s(q^2)-I_2^c(q^2)]\sin^2\theta_K]\cos 2\theta_\ell\nn\\[.4em] 
&+I_3(q^2)\sin^2\theta_K \sin^2\theta_\ell \cos 2\phi+I_4(q^2)\sin 2\theta_K \sin 2\theta_\ell \cos \phi \nn \\[.4em] 
&+ I_5(q^2) \sin 2\theta_K\sin \theta_\ell\cos\phi+[I_6^c(q^2)+[I_6^s(q^2)-I_6^c(q^2)]\sin^2\theta_K]\cos \theta_\ell \nonumber \\[.4em] 
&+I_7(q^2)\sin 2\theta_K \sin \theta_\ell \sin \phi + I_8(q^2)\sin 2\theta_K \sin 2\theta_\ell \sin\phi \nonumber\\[.4em] 
&+I_9(q^2) \sin^2\theta_K \sin^2\theta_\ell \sin 2 \phi.
\end{align}
After integrating over angles the differential decay rate is simply 
\begin{equation}
{\mathrm{d}{\cal B}\over \mathrm{d}q^2}=\frac{1}{4}\left[6 I_1^s(q^2)+3 I_1^c(q^2)-2I_2^s(q^2)-I_2^c(q^2)\right]\ .
\end{equation}
Similarly, the forward-backward asymmetry and the longitudinal polarization fraction of the lepton in the decay process are defined as follows:
\begin{equation}
\mathcal{A}_{FB}(q^2) = \frac{3 I_6^s(q^2) + \tfrac{3}{2} I_6^c(q^2)}{6 I_1^s(q^2)+ 3 I_1^c(q^2)- 2 I_2^s(q^2)  - I_2^c(q^2) },
\label{eq:AFB}
\end{equation}
\begin{equation}
\mathcal{F}_{L}(q^2) = \frac{3 I_1^c(q^2) - I_2^c(q^2)}{ 6 I_1^s(q^2) +3 I_1^c(q^2)  - 2 I_2^s(q^2)- I_2^c(q^2)}.
\label{eq:FL}
\end{equation}
The angular coefficients, which depend on the momentum transfer $q^2$, are constructed from combinations of the decay's helicity amplitudes. These amplitudes can also be represented using the transversity amplitudes $A_{\perp,\parallel,0,t}^{L(R)}\equiv A_{\perp,\parallel,0,t}^{L(R)}(q^2)$ in the following manner:
\begin{align}
A_{\perp}^{L(R)} &= {\cal N}_{K^\ast} \sqrt{2} \lambda_B^{1/2}\left[(C_9 \mp C_{10} )\frac{V(q^2)}{m_B+m_{K^\ast}}\right],\nn \\
A_{\parallel}^{L(R)} &=  -{\cal N}_{K^\ast} \sqrt{2}(m_B^2-m_{K^\ast}^2)\left[(C_9 \mp C_{10})\frac{A_1(q^2)}{m_B-m_{K^\ast}}\right],\nn  
\end{align}
\begin{align}
A_0^{L(R)}&=-\frac{{\cal N}_{K^\ast}}{2 m_{K^\ast} \sqrt{q^2}}(C_9 \mp C_{10})\left[ (m_B^2-m_{K^\ast}^2-q^2)(m_B+m_{K^\ast})A_1(q^2)-\frac{ \lambda_B A_2(q^2)}{m_B+m_{K^\ast}}\right],  \nonumber\\[0.7 em]
\label{eq:helicityamplitudest}
A_{t}^{L(R)} &=  -{\cal N}_{K^\ast} \frac{\lambda_B^{1/2}}{\sqrt{q^2}}\left[(C_9  \mp C_{10}) +\frac{q^2}{m_b+m_s}\left(\frac{C_S}{m_1-m_2}\mp \frac{C_P}{m_1+m_2}\right)\right] A_0(q^2),
\end{align}
where for shortness, $\lambda_B=\lambda(m_B,m_{K^\ast},\sqrt{q^2})$, $\lambda_q=\lambda(m_1,m_2,\sqrt{q^2})$, and 
\begin{equation}
{\cal N}_{K^\ast}=V_{tb}V_{ts}^\ast\left[ \frac{\tau_{B_d} G_F^2 \alpha^2}{3 \times 2^{10} \pi^5 m_B^3} \lambda_B^{1/2} \lambda_q^{1/2} \right]^{1/2}.
\end{equation}
The upper signs in the aforementioned equations pertain to $A_i^L$, while the lower signs relate to $A_i^R$. It's noteworthy that $A_{t}$ also carries the superscript $L(R)$, indicating the chirality of the lepton pair. This may appear unusual when juxtaposed with the lepton flavor-conserving case, and we will now provide an explanation for this.
 When $\ell_1=\ell_2$, the pseudoscalar density can be reformulated as follows,
\begin{equation}
\bar{\ell} \gamma_5 \ell = \frac{q^\mu}{2 m_\ell} (\bar{\ell} \gamma_\mu \gamma_5 \ell).
\end{equation}
This allows us to assimilate the contributions originating from the operator $\mathcal{O}_P^{(\prime)}$ into the amplitude $A_t$. This amplitude is associated with the time-like polarization vector of the virtual vector boson, denoted as $\epsilon_{V}^\mu(t)=q^\mu/\sqrt{q^2}$.
An analogous method is not applicable to the scalar operator $\mathcal{O}_S^{(\prime)}$ due to the fact that $q^\mu (\bar{\ell} \gamma_\mu \ell) = 0$. In this case, it is necessary to introduce a new amplitude, denoted as $A_S$, to account for the remaining scalar contribution. In LFV scenarios, where $m_1\neq m_2$, it becomes possible to utilize the Ward identities to absorb both the scalar and pseudoscalar densities into the vector and axial currents, respectively.
Consequently, in LFV cases, the amplitudes $A_t$ and $A_S$ are substituted by $A_t^{L(R)}$. Although the expressions for $A_t^{L,R}$ become problematic in the limit where $m_1 = m_2$, we have verified that the angular coefficients remain well-defined, and the standard formulas presented in reference \cite{Altmannshofer:2008dz} are recovered. Finally, the angular coefficients $I_{1-9}(q^2)$ in terms of the transversity amplitudes~(\ref{eq:helicityamplitudest}), are given by
\begin{align}
\label{eq:angular}
I_1^s(q^2) &=\frac{4 m_1 m_2}{q^2}\mathrm{Re}\left(A_{\parallel}^L A_{\parallel}^{R\ast}+A_{\perp}^L A_{\perp}^{R\ast}\right)+\biggl[|A_{\perp}^L|^2+|A_{\parallel}^L|^2+ (L\to R) \biggr]\frac{\lambda_q +2 [q^4-(m_1^2-m_2^2)^2]}{4 q^4}.  \\
I_1^c(q^2) &= \frac{8 m_1 m_2}{q^2} \mathrm{Re}(A_0^L A_0^{R\ast}-A_t^L A_t^{R\ast}) +\bigl[|A_0^L|^2+|A_0^R|^2 \bigr]\frac{q^4-(m_1^2-m_2^2)^2}{q^4}\nn\\
&\hspace{3.5cm}-2\frac{(m_1^2-m_2^2)^2-q^2 (m_1^2+m_2^2)}{q^4}\bigl(|A_t^L|^2+|A_t^R|^2\bigr).\\ \nn
\end{align}
\begin{align}
I_2^s(q^2) &= \frac{\lambda_q}{4 q^4}[|A_\parallel^L|^2+|A_\perp^L|^2+(L\to R)].\\
I_2^c(q^2) &= - \frac{\lambda_q}{q^4}(|A_0^R|^2+|A_0^L|^2).  \\
I_3(q^2) &= \frac{\lambda_q}{2 q^4} [|A_\perp^L|^2-|A_\parallel^L|^2+(L\to R)].\\
I_4(q^2) &= - \frac{\lambda_q}{\sqrt{2} q^4} \mathrm{Re}(A_\parallel^L A_0^{L\ast}+(L\to R)]. \\
I_5(q^2) &= \frac{\sqrt{2}\lambda_q^{1/2}}{q^2} \left[ \mathrm{Re}(A_0^L A_\perp^{L\ast}-(L\to R)) -\frac{m_1^2-m_2^2}{q^2} \mathrm{Re}(A_t^L A_\parallel^{L\ast}+(L\to R))\right]. \\
I_6^s(q^2) &=- \frac{2 \lambda_q^{1/2}}{q^2}[\mathrm{Re}(A_\parallel^L A_\perp^{L\ast}-(L\to R))]. \\
I_6^c(q^2) &= - \frac{4\lambda_q^{1/2}}{q^2}\frac{m_1^2-m_2^2}{q^2} \mathrm{Re}(A_0^L A_t^{L\ast}+(L\to R) .\\
I_7(q^2) &= - \frac{\sqrt{2}\lambda_q^{1/2}}{q^2} \left[ \frac{m_1^2-m_2^2}{q^2} \mathrm{Im}(A_\perp^{L}A_t^{L\ast} +(L\to R))+ \mathrm{Im}(A_0^L A_\parallel^{L\ast}-(L\to R))\right]. \\
I_8(q^2) &= \frac{\lambda_q}{\sqrt{2}q^4}\mathrm{Im}(A_0^{L}A_\perp^{L\ast} +(L\to R)). \\
I_9(q^2) &=- \frac{\lambda_q}{q^4}\mathrm{Im}(A_\perp^R A_\parallel^{R\ast}+A_\perp^L A_\parallel^{L\ast}  ).
\end{align}
The form factors associated with the transversity amplitude are obtained using the LCSR method  \cite{Bharucha:2015bzk} and are expressed as follows:
\begin{equation}
    F_{y}(q^{2})=\frac{1}{1 - q^{2}/ m_{R,y}^{2}} \sum_{k=0,1,2}\beta_{k}[z(q^{2})-z(0)]^{k},
\end{equation}
where $z(q^{2})=\frac{\sqrt{t_{+}-q^{2}}- \sqrt{t_{+}-t_{0}}}{\sqrt{t_{+}-q^{2}}+ \sqrt{t_{+}-t_{0}}}$,~ $t_{\pm}= (m_{B}{\pm} m_{K^{*}})^2$, ~$t_{0}=t_{+}(1-\sqrt{1-t_{-}/t_{+}})$ and $m_{R,F_y}$ is the resonance mass corresponding to the form factor. In our calculations, we use the resonance masses as follows: 
\begin{eqnarray}
m_{R,(V,T_1)}=5.412 \hspace{0.1cm}{\rm GeV},  \hspace{0.5cm} m_{R,A_{0}}=5.336  \hspace{0.1cm}{\rm GeV}, \hspace{0.5cm} m_{R,(T_2,T_3,A_1,A_3)}=5.829 \hspace{0.1cm}{\rm GeV}.
\end{eqnarray}
 We apply the above formalism originally developed for $B \to K^{*} $ decay to analyze the process $B \to \phi \ell_1 \ell_2$. This is achieved by straightforwardly substituting the relevant mass and form factors for the corresponding vector meson $\phi$. The numerical values for the parameters associated with the form factors with $1\sigma$ uncertainty are given in Table \ref{tab:Formfactors for k}.
\begin{table}[h!]
\centering
\begin{tabular}{||c|c|c|c||}
\hline\hline

$F_{i}$ & $\beta_{0}$ & $\beta_{1}$ &$ \beta_{2}$ \\
\hline
$V$ & $0.34 \pm 0.04$ & $-1.05\pm 0.24$& $2.37\pm 1.39$\\
$ T_1$ & $0.28\pm0.03$ &$ -0.89\pm0.19$ &4$1.95\pm1.10$ \\
$ T_2$ &$0.28\pm0.03$& $0.40\pm0.18$ &$0.36\pm0.51$ \\
$T_3$ &$0.67\pm0.08$ & $1.48\pm0.49$ &$1.92\pm1.96$ \\
$ A_{0}$ &$0.36\pm 0.05$ &$-1.04\pm 0.27$&$1.12\pm1.35$ \\
$A_1$ &$0.27\pm0.03$ & $0.39\pm0.19$ &$-0.11\pm0.48$ \\
$ A_3$ &$0.26\pm0.03$ & $0.60\pm0.20$ &$0.12\pm0.84$ \\

\hline \hline
\end{tabular}
\hspace{7mm}
\begin{tabular}{||c|c|c|c||}
\hline\hline
$F_{i}$ & $\beta_{0}$ & $\beta_{1}$ &$ \beta_{2}$ \\
\hline
$V$ & $0.39 \pm 0.03$ & $-1.03\pm0.25$& $3.50\pm1.55$\\
$ A_{0}$ &$0.39\pm0.05$ &$-0.78\pm0.26$&$2.41\pm1.48$ \\
$A_1$ &$0.30\pm0.03$ & $0.48\pm0.19$ &$0.29\pm0.65$ \\
$ A_3$ &$0.25\pm0.03$ & $0.76\pm0.20$ &$0.71\pm0.96$ \\
$ T_1$ & $0.31\pm0.03$ &$ -0.87\pm0.19$ &4$2.75\pm1.19$ \\
$ T_2$ &$0.31\pm0.03$& $0.58\pm0.19$ &$0.89\pm0.71$ \\
$T_3$ &$0.68\pm0.07$ & $2.11\pm0.46$ &$4.94\pm2.25$ \\
\hline \hline
\end{tabular}
\caption{Form factors for $B \to K^{*}\ell_1 \ell_2$ (left panel) and $B_{s} \to \phi \ell_1 \ell_2$ (right panel).}
\label{tab:Formfactors for k}
\end{table} 
\subsection*{Numerical analysis for  $B\to K^*(\to K\pi) \ell_1 \ell_2$ and $B_s\to \phi(\to K\bar K) \ell_1 \ell_2$}
To quantitatively highlight the significance of the terms that scale the Wilson coefficients, we utilize the form factors provided in Ref.~\cite{Ball:2004rg}, and distinguish between scenarios where LFV originates from vector-type and scalar-type operators. The scaling factors (multipliers) for the Wilson coefficients are obtained by integrating over the full kinematic range of the dilepton invariant mass squared, $q^2$. It is noteworthy that the integrand functions exhibit distinct behaviors: contributions from terms proportional to $|C_{9,10}|^2$ dominate in the intermediate $q^2$ region, while those involving $|C_{S,P}|^2$ become more significant at higher values of $q^2$. To illustrate this behavior, we analyze the decay channels $\tau^+ \mu^-$ and $\tau^- \mu^+$ separately.

\begin{figure}[h!]
\begin{tabular}{ccc}
	\centering
	\includegraphics[width=0.33\textwidth]{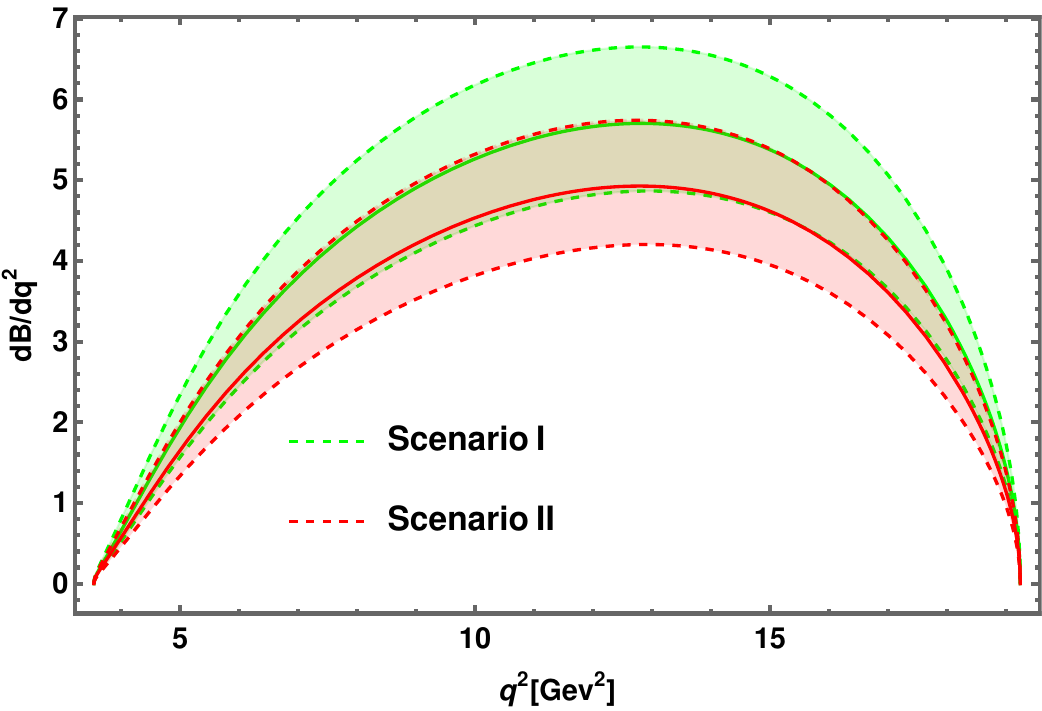} &
  	\includegraphics[width=0.33\textwidth]{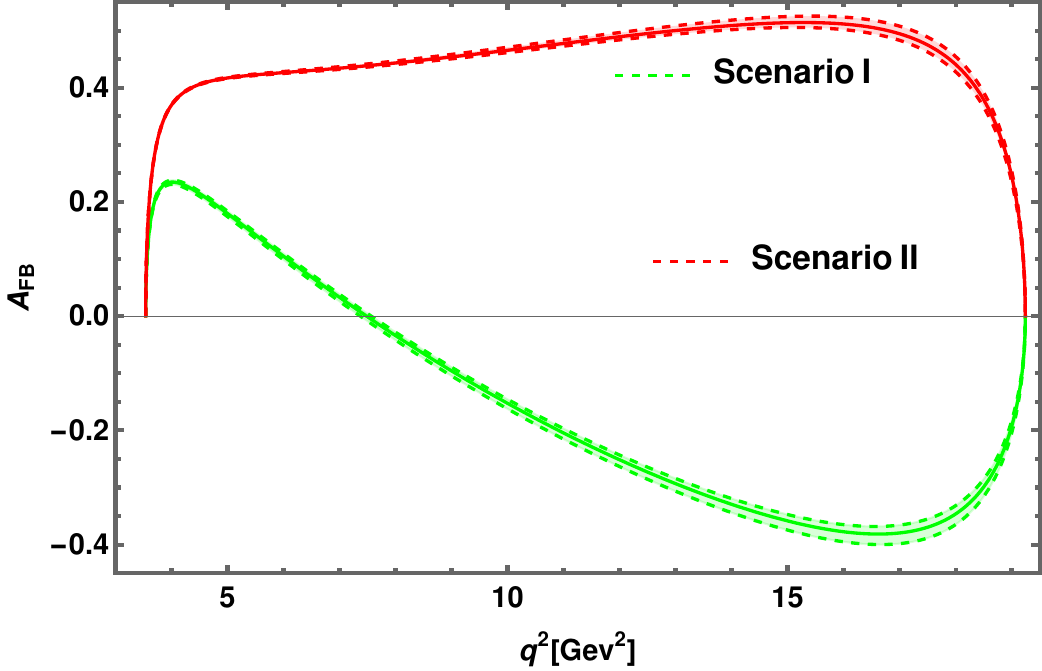}&
  	\includegraphics[width=0.33\textwidth]{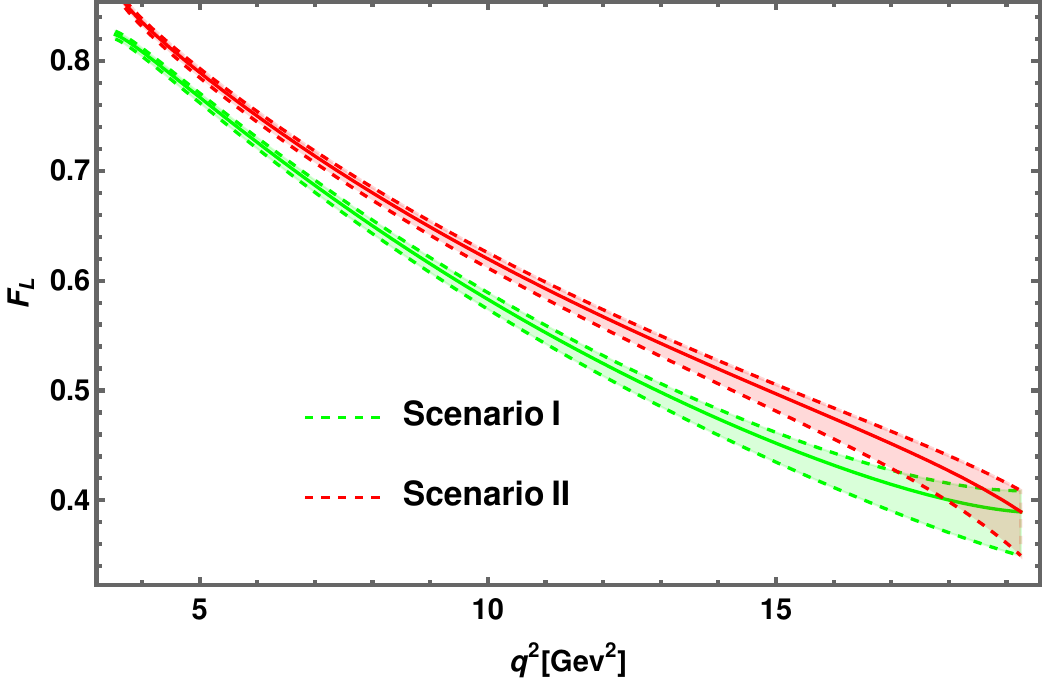} \\
  	\end{tabular}
    \caption{ Branching ratio (in units of $10^{-5}$)  (left), forward-backward asymmetry (middle) and lepton polarization (right) for $B\to K^* \tau^+\mu^-$.}
	\label{fig::kstarmumtaup}
\end{figure}

\begin{figure}[h!]
\begin{tabular}{ccc}
	\centering
	\includegraphics[width=0.33\textwidth]{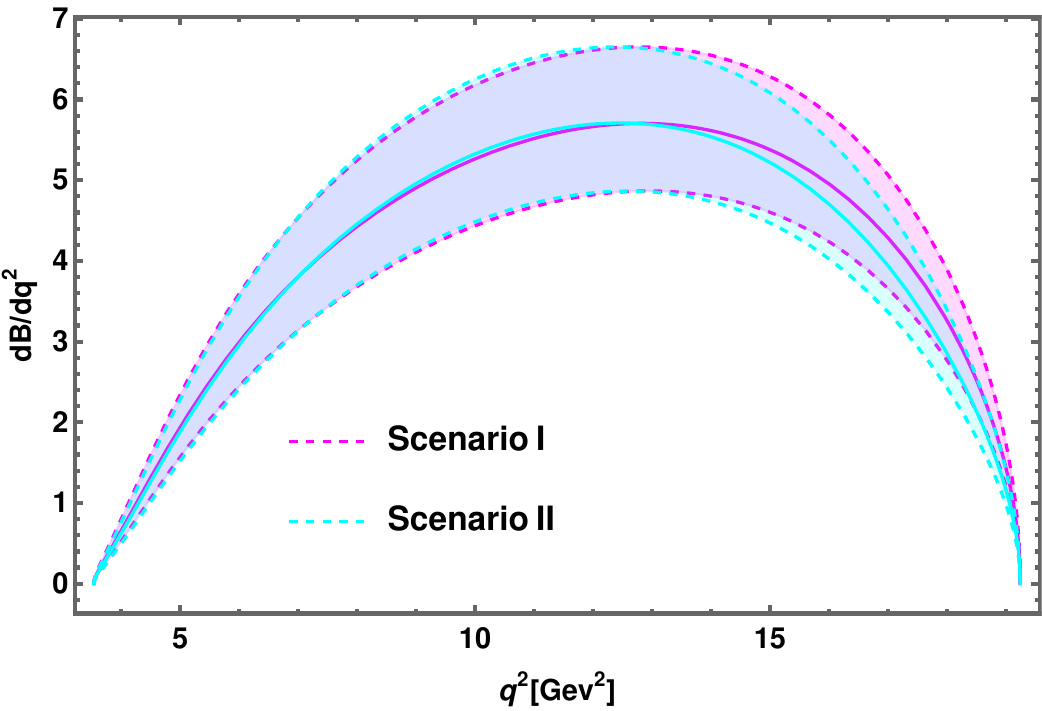} &
  	\includegraphics[width=0.33\textwidth]{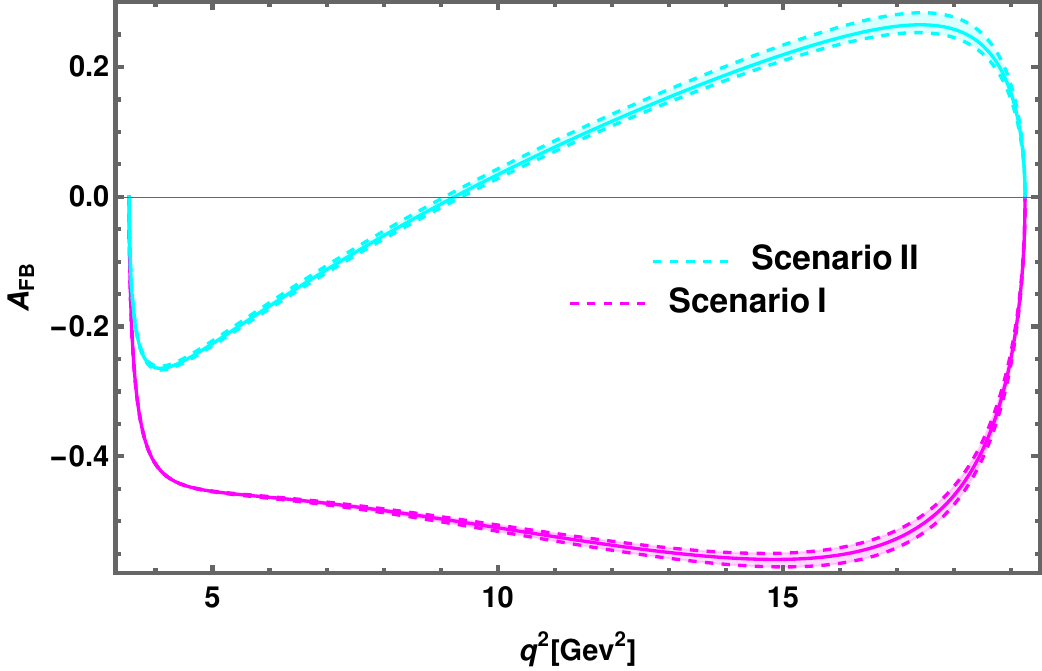}&
  	\includegraphics[width=0.33\textwidth]{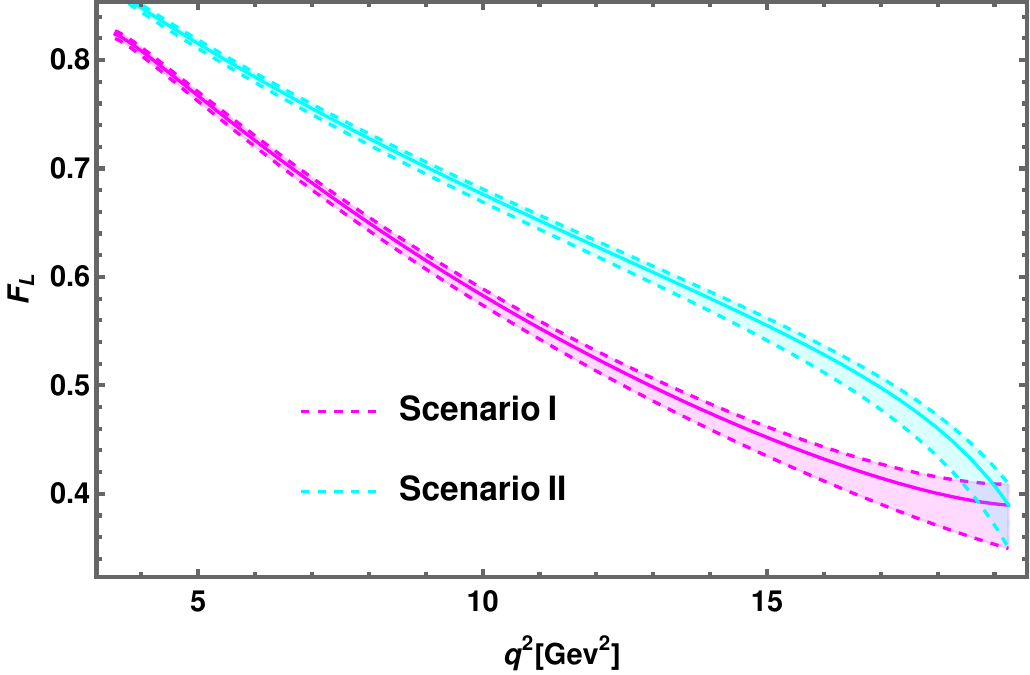} \\
  	\end{tabular}
    \caption{  Branching ratio (in units of $10^{-5}$) (left), forward-backward asymmetry (middle) and lepton polarization (right) for $B\to K^* \tau^-\mu^+$.}
	\label{fig::kstarmuptaum}
\end{figure}

\begin{figure}[h!]
\begin{tabular}{ccc}
	\centering
	\includegraphics[width=0.33\textwidth]{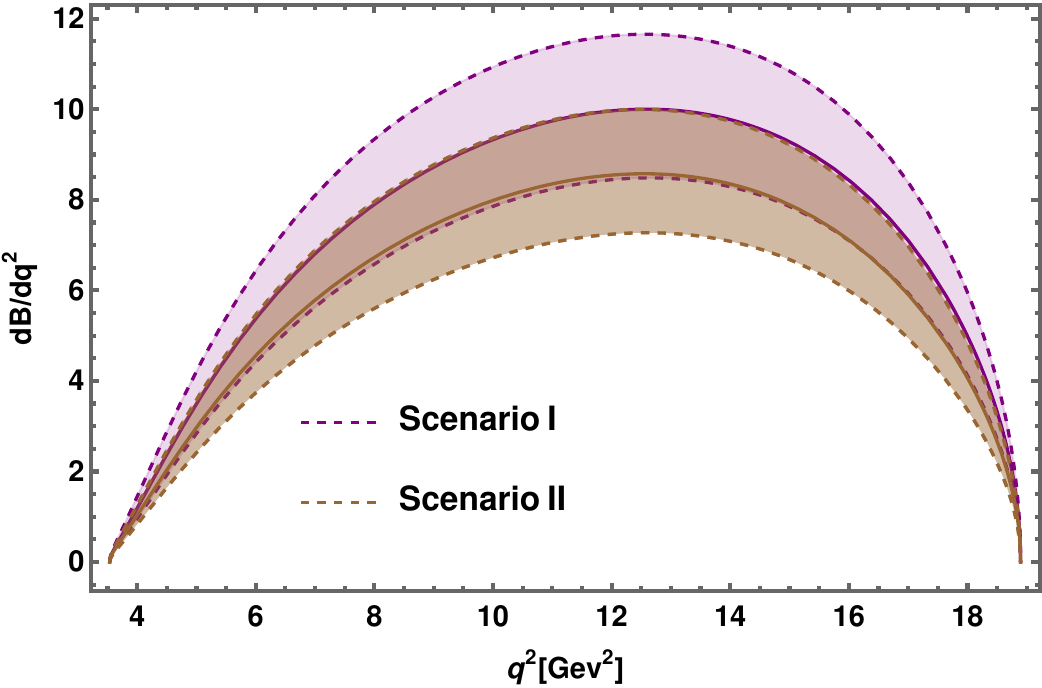} &
  	\includegraphics[width=0.33\textwidth]{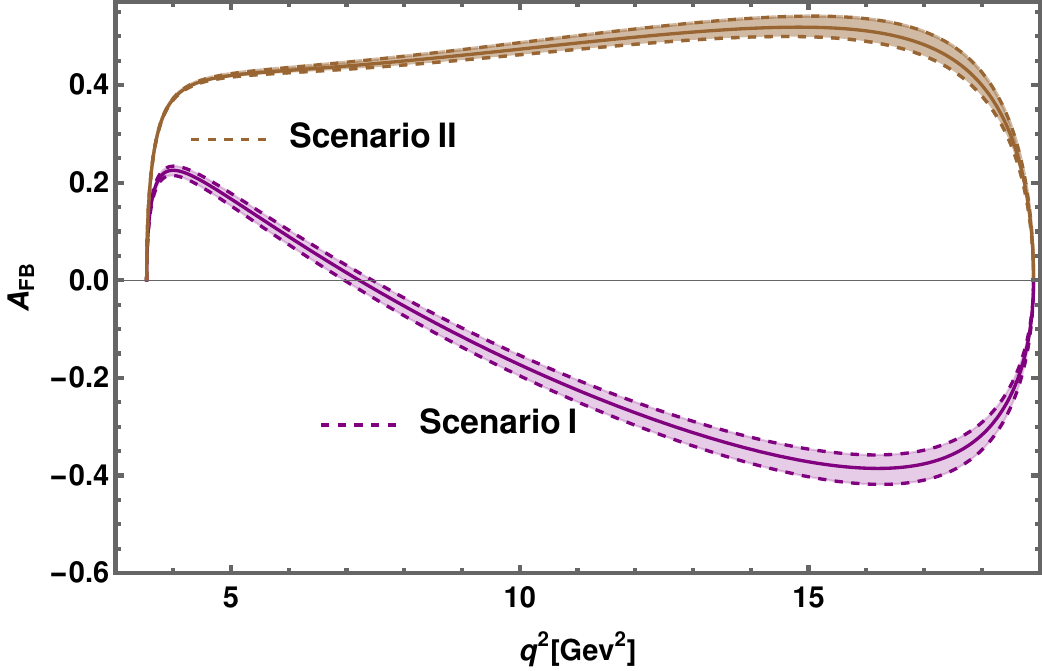}&
  	\includegraphics[width=0.33\textwidth]{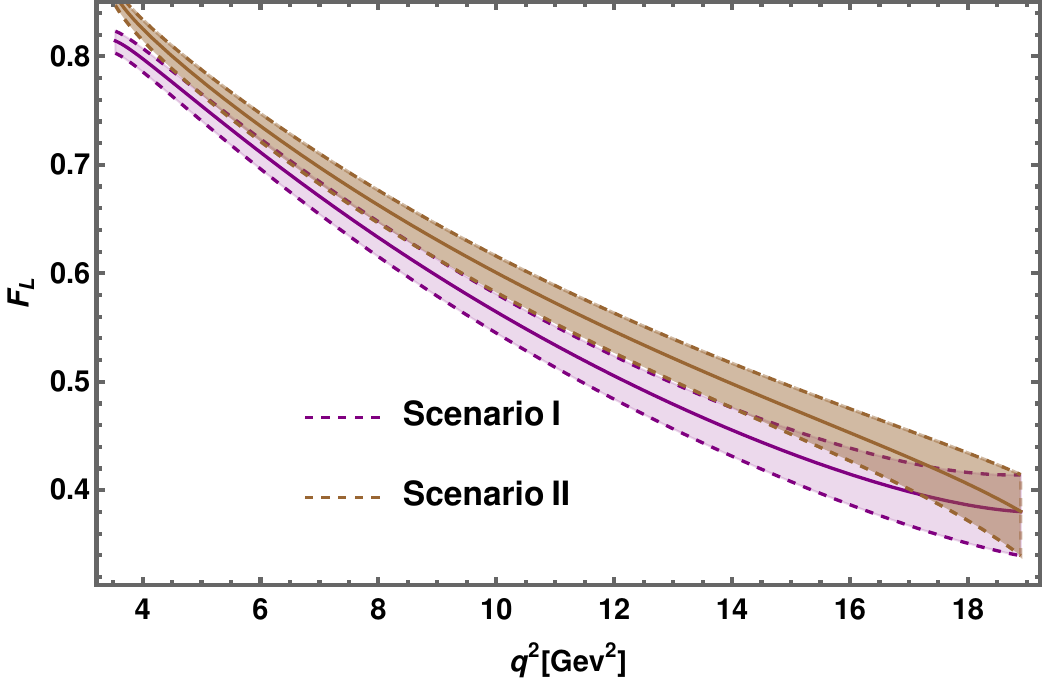} \\
  	\end{tabular}
    \caption{  Branching ratio (in units of $10^{-5}$) (left), forward-backward asymmetry (middle) and lepton polarization (right) for $B\to \phi \tau^+\mu^-$.}
	\label{fig::phimumtaup}
\end{figure}

\begin{figure}[h!]
\begin{tabular}{ccc}
	\centering
	\includegraphics[width=0.33\textwidth]{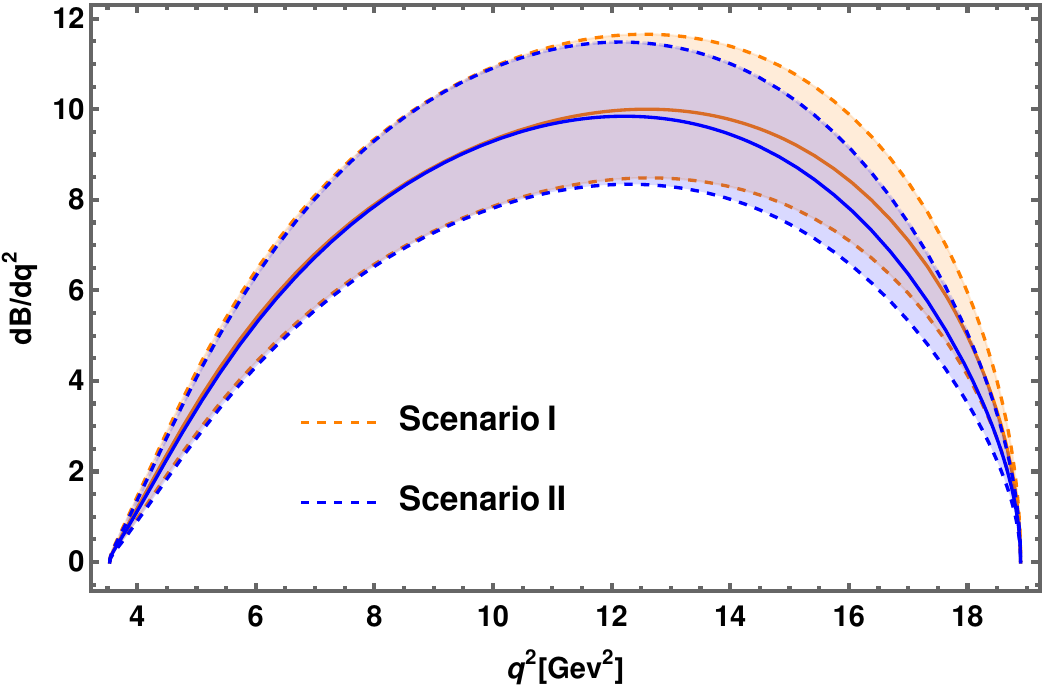} &
  	\includegraphics[width=0.33\textwidth]{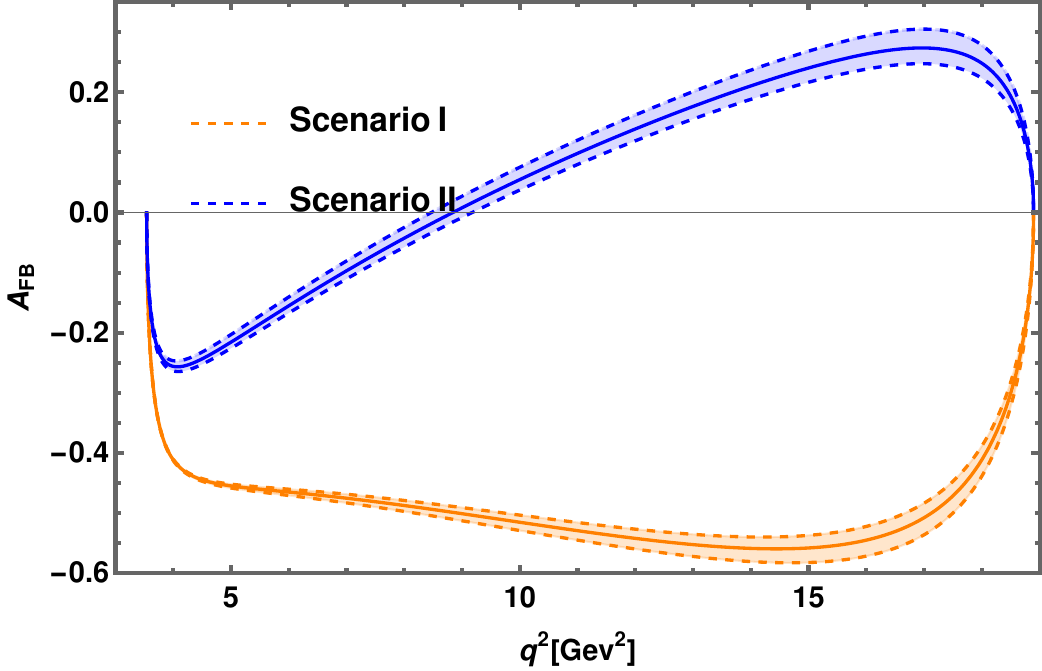}&
  	\includegraphics[width=0.33\textwidth]{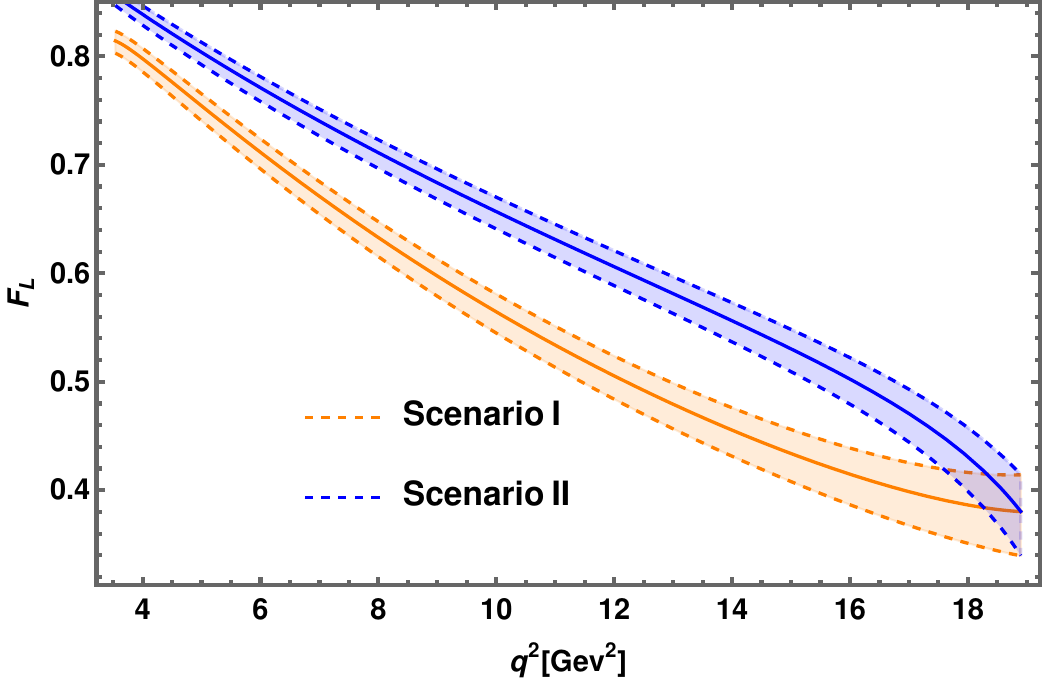} \\
  	\end{tabular}
    \caption{  Branching ratio (in units of $10^{-5}$) (left), forward-backward asymmetry (middle) and lepton polarization (right) for $B\to \phi \tau^-\mu^+$.}
	\label{fig::phimuptaum}
\end{figure}

In Figure~\ref{fig::kstarmumtaup}, we present the differential distributions for the $B \to K^* \tau^+ \mu^-$ decay. The plots display the $\mathcal{B}$ (left), $\mathcal{A}_{FB}$(middle), and $\mathcal{F}_L$ (right), highlighting the differences between Scenario~I (green ) and Scenario~II (red ), each shown with their corresponding $1\sigma$ uncertainty bands. Similarly, Figure~\ref{fig::kstarmuptaum} shows the corresponding observables for the $B \to K^* \tau^- \mu^+$ decay, with Scenario~I illustrated in magenta and Scenario~II in cyan blue, again including $1\sigma$ error bands. In Figure~\ref{fig::phimumtaup}, we present the differential distributions for the $B \to \phi \tau^+ \mu^-$ decay. The plots display the $\mathcal{B}$ (left), $\mathcal{A}_{FB}$(middle), and $\mathcal{F}_L$ (right), highlighting the differences between Scenario~I (purple ) and Scenario~II (brown ), each shown with their corresponding $1\sigma$ uncertainty bands. Similarly, Figure~\ref{fig::phimuptaum} shows the corresponding observables for the $B \to \phi \tau^- \mu^+$ decay, with Scenario~I illustrated in orange Scenario~II in blue, again including $1\sigma$ error bands.

Here, in the branching ratio plots, we can see that each mode of the decay process has almost similar $q^2$ dependencies irrespective of the charge of the heavier lepton, but in case of $\tau^+\mu^-$ channel the central value of scenario I is higher than the scenario II. $\mathcal{A}_{FB}$ is particularly sensitive to the chiral structure of the underlying NP interactions. A consistent zero-crossing point near $q^2 \approx 9 \, \text{GeV}^2$ appears in both plots, characteristic of the interplay between vector and axial-vector contributions in the decay dynamics. Scenario~I demonstrates a significant positive asymmetry for $\tau^+\mu^-$ channel at higher $q^2$ values (peaking near $0.4$), while Scenario~II of $\tau^-\mu^+$  remains predominantly negative. This divergence suggests potential new physics contributions. $\mathcal{F}_{L}$ in \(B \rightarrow K^* \ell^+ \ell^-\) decays exhibits consistent behavior across Scenarios~I and II, with values spanning \(0.4 \leq F_L \leq 0.8\). Both scenarios show a monotonic decrease in \(F_L\) with increasing \(q^2\), suggesting a growing contribution from transverse amplitudes at higher dilepton invariant masses. The negligible difference between \(\mu\) and \(\tau\) final states indicates that \(F_L\) is less sensitive to lepton mass effects compared to observables like \(A_{FB}\). Marginal deviations between scenarios at high \(q^2\) (\(\sim 15 \, \text{GeV}^2\)) may reflect theoretical uncertainties or new physics contributions to the helicity structure. Finally, Table~\ref{tab:resultskstar} summarizes the numerical values of the $\mathcal{B}$, $\mathcal{A}_{FB}$, and $\mathcal{F}_L$ for the $B \to K^* \tau^\pm \mu^\mp$ LFV decays. These results incorporate the upper limit constraints at $90\%$ confidence level for both scenarios.

\begin{table}[h!]
\begin{center}
\renewcommand{\arraystretch}{1.2} 
\begin{tabular}{||c| c| c| c||}
\hline\hline
\textbf{Decay modes}& \textbf{Observables} & \textbf{Scenario I}  & \textbf{Scenario II} \\
 \hline
 & $ \mathcal{B}$ & $ <6.547 \times 10^{-5}$  & $ <5.606\times 10^{-5}$ \\
 $B \to K^* \tau^+\mu^- $ &$\mathcal{A_{FB}}$ & $< -0.156$ & $< 0.045$ \\
 & $\mathcal{F_L}$& $<0.563$  & $ <0.596$ \\
\hline\hline
& $ \mathcal{B}$ & $ <6.514 \times 10^{-5}$  & $ <6.415 \times 10^{-5}$ \\
$B \to K^* \tau^-\mu^+ $ &$\mathcal{A_{FB}}$ & $< -0.490$ & $< 0.055$ \\
 & $\mathcal{F_L}$& $<0.561$  & $ <0.642$ \\
\hline\hline
 & $ \mathcal{B}$ & $ <1.127 \times 10^{-5}$  & $ <1.577 \times 10^{-5}$ \\
 $B \to \phi \tau^+\mu^- $ &$\mathcal{A_{FB}}$ & $< -0.164$ & $< 0.455$ \\
 & $\mathcal{F_L}$& $<0.550$  & $ <0.583$ \\
\hline\hline
& $ \mathcal{B}$ & $ <1.123 \times 10^{-5}$  & $ <1.087 \times 10^{-5}$ \\
$B \to \phi \tau^-\mu^+ $ &$\mathcal{A_{FB}}$ & $< -0.491$ & $< 0.066$ \\
 & $\mathcal{F_L}$& $<0.557$  & $ <0.627$ \\
\hline\hline
\end{tabular}
\caption{Upper limits for the branching ratio, forward-backward asymmetry($\mathcal{A_{FB}}$) and lepton polarization ($\mathcal{F_L}$) of $B \to (K^*, \phi)\tau^{\pm}\mu^{\mp}$ LFV decays considering bounds are at $90\%$ C.L..}
\label{tab:resultskstar}
\end{center}
\end{table}
\section{Decay distribution of $B\to K^*_2\ell_1\ell_2$\label{sec:k2star}}
The NP Wilson coefficients $C^{(\prime)}_{V, A, S, P}$ encompass the short-distance physical aspects, while the long-distance characteristics are integrated into the hadronic matrix elements of $B\to K_2^*$. In the context of the $B\to K_2^*$ transition, the hadronic matrix elements associated with the V and A currents can be described using four form factors: $V(q^2)$ and $A_{0,1,2}(q^2)$. These can be written as~\cite{Wang:2010ni}
 \begin{eqnarray}
 \langle  K_2^*(k,\epsilon^*)|\bar s\gamma^{\mu}\gamma_5 b|\overline B(p)\rangle
   &=&i(m_B+m_{K_2^*})A_1(q^2)\left[ \epsilon^{*\mu}_{T}
    -\frac{\epsilon^*_{T } \cdot  q }{q^2}q^{\mu} \right]+2im_{K_2^*} A_0(q^2)\frac{\epsilon^*_{T } \cdot  q }{ q^2}q^{\mu}  \nonumber\\
    &&-iA_2(q^2)\frac{\epsilon^*_{T} \cdot  q }{  m_B+m_{K_2^*} }
     \left[ (p+k)^{\mu}-\frac{m_B^2-m_{K_2^*}^2}{q^2}q^{\mu} \right]
 , \nonumber\\
  \langle K_2^*(k, \epsilon^*)|\bar s\gamma^{\mu}b|\overline B(p)\rangle
  &=&-\frac{2V(q^2)}{m_B+m_{K_2^*}}\epsilon^{\mu\nu\rho\sigma} \epsilon^*_{T\nu}  p_{\rho}k_{\sigma} .\end{eqnarray}
However, by applying the equation of motion, it can be demonstrated that the matrix element for the $B\to K^*_2$ transition involving the scalar interaction ($\bar{s}b$) becomes null, while the one associated with the pseudo-scalar interaction can be expressed as:
\begin{equation}
 \langle K_2^*(k, \epsilon^*)|\bar s\gamma_{5}b|\overline B(p)\rangle =  -\frac{2i m_{K^*_2} A_0(q^2)}{m_b + m_s} (\epsilon^*_{T} \cdot  q).
\end{equation}
Here, $p$ and $k$ represent the four-momentum of the $B$ and $K_2^*$ mesons, respectively.
The details regarding the polarization vector of $K^*_2$ can be found in Appendix~\ref{appenb}. All the form factors have been computed using both the perturbative QCD approach \cite{Wang:2010ni} and the Light-Cone QCD sum rule technique \cite{Aliev:2019ojc,Wang:2010tz}. For this study, we have adopted the most up-to-date values of the form factors, as determined through the Light-Cone QCD sum rule (LCSR) method, as presented in Ref.~\cite{Aliev:2019ojc}. Within this method, each form factor can be expressed as follows: 
\begin{equation}
\label{eqn:FormFactor}
F^{B\to K_2^*}(q^2)=\frac{1}{1-q^2/m^2_{R,F}}\sum_{n=0}^{1}\beta_n^F\left[z(q^2)-z(0)\right]^n,
\end{equation}
where $z(q^2)=\frac{\sqrt{t_+-s}- \sqrt{t_+-t_0}}{\sqrt{t_+-s}+ \sqrt{t_+-t_0}}$, $t_{\pm}=(m_B\pm m_{K_2^*})^2$, $t_0=t_+(1-\sqrt{1-t_-/t_+})$ and $m_{R,F}$ is the resonance mass associated with the quantum numbers of corresponding form factor. The resonance masses employed in our numerical calculations are given as
\begin{eqnarray}
m_{R,V}=5.412 \hspace{0.1cm}{\rm GeV}, \hspace{0.5cm} m_{R,A_{0}}=5.336 \hspace{0.1cm}{\rm GeV}, \hspace{0.5cm} m_{R,(A_1,A_3)}=5.829 \hspace{0.1cm}{\rm GeV}.
\end{eqnarray}. 
The fit parameters $\beta_n^F$ are given in Table~\ref{tab:FormFactor}.
\begin{table}[h!]
\begin{tabular}{||c|c|c||}
\hline \hline
Form Factor &  $\beta_1$ &$\beta_0$\\
\hline
$V^{B\to K_2^*}$ & $-0.90^{+0.37}_{-0.50}$ &$0.22^{+0.11}_{-0.08}$\\
\hline
$A_0^{B\to K_2^*}$ & $-1.23^{+0.23}_{-0.23}$  & $0.30^{+0.01}_{-0.05}$\\
\hline
$A_1^{B\to K_2^*}$ & $-0.46^{+0.19}_{-0.25}$ & $0.19^{+0.09}_{-0.07}$ \\
\hline
$A_2^{B\to K_2^*}$ & $-0.40^{+0.23}_{-0.16}$ & $0.11^{+0.05}_{-0.06}$ \\
\hline \hline
\end{tabular}
\caption{Fits for $B\to K_2^*$ form factors using LCSR approach.}
\label{tab:FormFactor}
\end{table}
The three-body $B \to K_2^* \ell_1 \ell_2$ decay can be characterized using the leptonic polar angle $\theta_{\ell}$ and the squared leptonic mass, denoted as $q^2 = (p-k)^2$. We define the $\theta_{\ell}$ angle as the angle formed by the $\ell_1$ lepton concerning the rest frame of the dilepton system. In terms of these two variables, the two-fold differential decay distribution can be expressed as follows: 
\begin{equation}
    \frac{d^2\Gamma}{dq^2 d \cos \theta_\ell}=A(q^2) + B(q^2) \cos \theta_\ell + C(q^2) \cos ^2\theta_\ell,
    \label{dist}
\end{equation}
where
\begin{eqnarray}
C(q^2)&=&\frac{3}{8}\bmi^2 \bpl^2   \left\lbrace\left(|A_L^{\parallel}|^2+|A_L^{\perp}|^2-2|A_L^{0}|^2\right)+\left(L\to R\right)\right\rbrace,\\\label{dist:C}
B(q^2)&=&\frac{3}{2}\bpl \bmi \left\lbrace \re\left[A_{L}^{\perp *} A_L^{\parallel} -(L \to R)\right]+ \frac{m_+m_-}{q^2} \re\left[A_{L}^{0*} A_L^{t} +(L \to R)\right]\right.\nn\\
&& \left. + \frac{m_{+}}{\sqrt{\qsq}} \re\left[A_S^* (A_L^0 +A_R^0) \right]-\frac{\mm} {\sqrt{\qsq}} \re\left[A_{SP}^* (A_L^0 -A_R^0) \right]\right\rbrace,\\
A(q^2) &=& \frac{3}{4}\left\lbrace\frac{1}{4}\left[\left(1+\frac{\mm^2}{q^2}\right)\bpl^2+\left(1+\frac{\mpl^2}{q^2}\right)\bmi^2 \right]\left(|A_L^{\parallel}|^2+|A_L^{\perp}|^2+(L\to R)\right)\right.\nn\\
&&+\frac{1}{2}\left(\bpl^2 +\bmi^2\right)\left(|A_L^{0}|^2+|A_R^{0}|^2\right)\nn\\
&&+\frac{4m_1 m_2}{q^2} \re\left[ A_L^{0*} A_R^0 + A_L^{\parallel *}A_R^{\parallel}+A_R^{\perp} A_L^{\perp*}-A_L^t A_R^{t*}\right]\nn\\
&&+\frac{1}{2}\left(\bpl^2+\bmi^2-2\bmi^2\bpl^2\right)\left(|A_L^t|^2 + |A_R^t|^2\right) +\frac{1}{2} \left(|A_{SP}|^2 \bmi^2 +|A_{S}|^2 \bpl^2\right)\nn\\
&&\left.+\frac{2\mm}{ \sqrt{q^2}}\bpl^2 \re\left[A_S(A_L^t+A_R^t)^*\right]- \frac{2\mpl}{\sqrt{q^2}} \bmi^2 \re\left[A_{SP}(A_L^t-A_R^t)^*\right]\right\rbrace\label{dist:A}.
 \end{eqnarray}
Here $m_{\pm}=(m_1 \pm m_2)$, $\beta_{\pm} = \sqrt{1-\frac{(m_{\ell_{1}} \pm m_{\ell_{2}})^2}{q^2}} $ and $A$'s are the transversity amplitudes. Integrating Eq.~(\ref{dist}) over $\theta_\ell$, we get the differential decay rate
\begin{equation}
    \frac{d\Gamma}{d q^2}= 2\left(A  + \frac{C}{3}\right)
\end{equation}
whereas the Lepton forward-backward asymmetry is found to be
\begin{equation}
A_{\rm FB}(q^2)= \frac{1}{d\Gamma/dq^2}\left(\int_0^1 d\cos\theta_\ell\frac{d^2\Gamma}{d\cos\theta_\ell d q^2}-\int_{-1}^0 d\cos\theta_\ell \frac{d^2\Gamma}{d\cos\theta_\ell d q^2 }\right) = \frac{B}{2\left(A+C/3\right)}.
\end{equation}
\subsection*{Numerical Analysis for  $B\to K^*_2\ell_1 \ell_2$}  
\begin{figure}[htb]
\begin{tabular}{cc}
	\centering
	\includegraphics[width=0.48\textwidth]{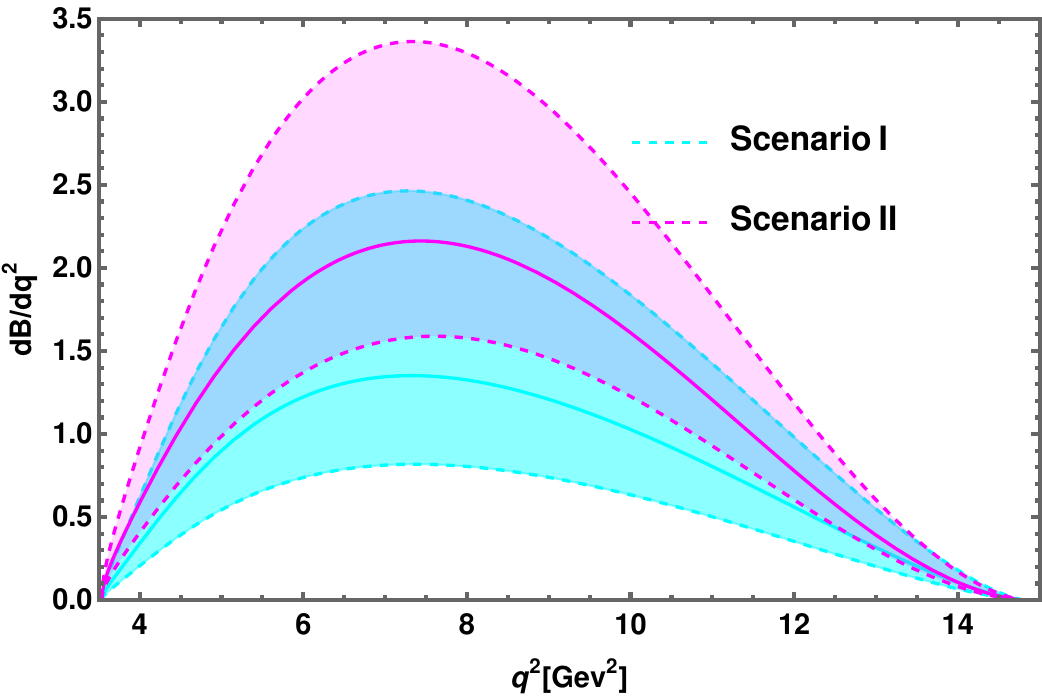} &
  	\includegraphics[width=0.48\textwidth]{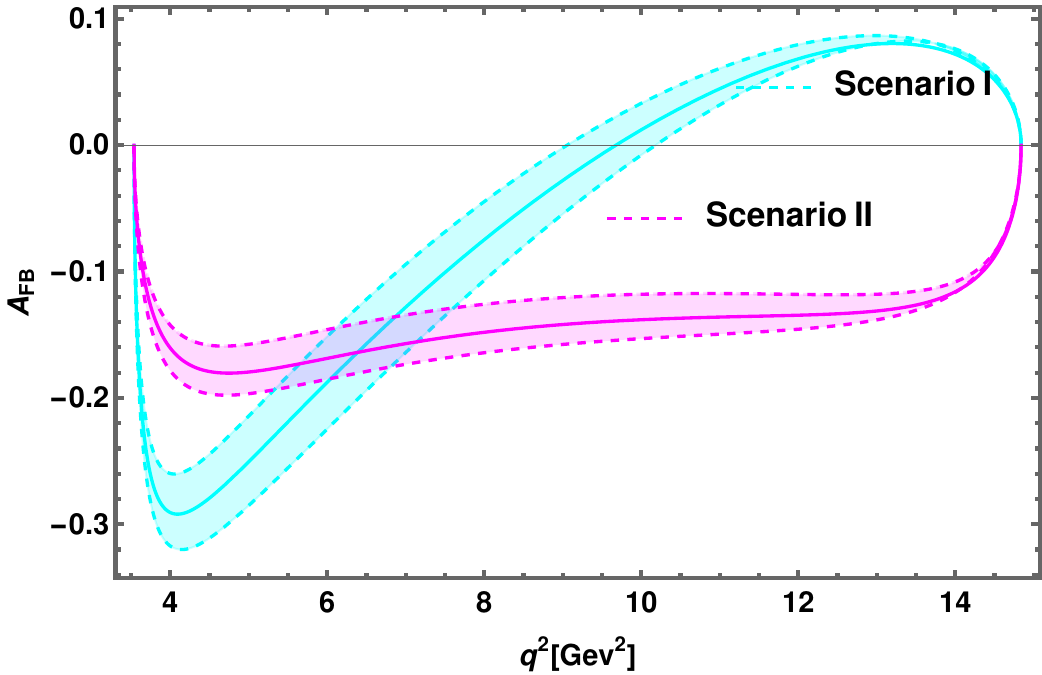}\\
  	\end{tabular}
    \caption{ Branching ratio (in units of $10^{-6}$) (left) and forward-backward asymmetry (right) for $B\to K_2^* \tau^+\mu^-$.}\label{fig::k2startaupmum}
    \end{figure}

\begin{figure}[htb]
\begin{tabular}{cc}
	\centering
	\includegraphics[width=0.48\textwidth]{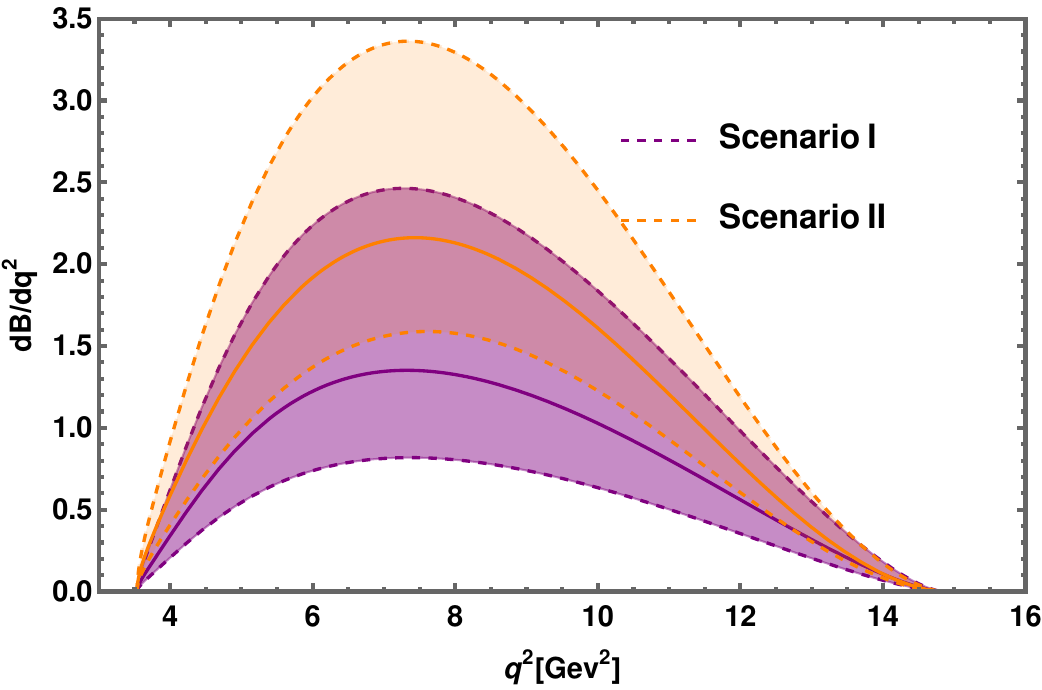} &
  	\includegraphics[width=0.48\textwidth]{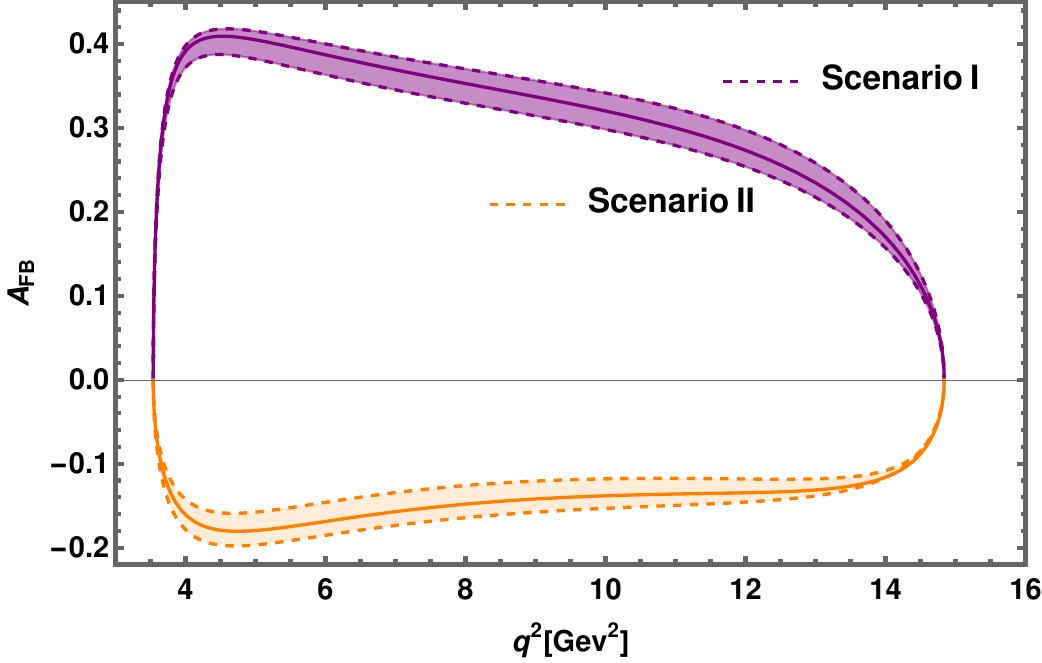}\\
  	\end{tabular}
    \caption{ Branching ratio (in units of $10^{-6}$) (left) and forward-backward asymmetry (right) for $B\to K_2^* \tau^-\mu^+$.}\label{fig::k2startaummup}
    \end{figure}

In Figure~\ref{fig::k2startaupmum}, we present the differential distributions for the $B \to K^*_2 \tau^+ \mu^-$ decay. The plots display the $\mathcal{B}$ (left) and $\mathcal{A}_{FB}$(right) highlighting the differences between Scenario~I (cyan blue ) and Scenario~II (magenta), each shown with their corresponding $1\sigma$ uncertainty bands. Similarly, Figure~\ref{fig::k2startaummup} shows the corresponding observables for the $B \to K^*_2 \tau^- \mu^+$ decay, with Scenario~I illustrated in purple and Scenario~II in orange, again including $1\sigma$ error bands. 

The branching ratios for the decays $B \rightarrow K_2^* \ell^+ \ell^-$ exhibit qualitatively similar behavior for both $\tau^\pm \mu^\mp$ channels. In both scenarios, a pronounced peak appears in the low-$q^2$ region, followed by a steep decrease. Scenario~II yields a higher central value compared to Scenario~I. The differential branching ratio plots indicate that the overall magnitude remains consistent between both decay modes. Regarding the forward-backward asymmetry $\mathcal{A}_{FB}$, Scenario~II predicts negative values throughout the full $q^2$ spectrum for both processes. In contrast, Scenario~I shows a zero-crossing at $q^2 = 10~\text{GeV}^2$ for the $B \to K^*_2 \tau^+ \mu^-$ channel. Table~\ref{tab:resultsk2star} summarizes the numerical results for both the branching ratios and forward-backward asymmetries in the lepton flavor violating decays $B \to K^*_2 \tau^+ \mu^-$ and $B \to K^*_2 \tau^- \mu^+$, evaluated under the assumption of 90\% confidence level upper limit constraints.

  \begin{table}[h!]
\begin{center}
\renewcommand{\arraystretch}{1.2} 
\begin{tabular}{||c| c| c| c||}
\hline\hline
\textbf{Decay modes}& \textbf{Observables} & \textbf{Scenario I}  & \textbf{Scenario II} \\
 \hline
$B \to K^*_2 \tau^+\mu^- $ & $ \mathcal{B}$ & $ <9.131 \times 10^{-6}$  & $ <1.416\times 10^{-6}$ \\
  &$\mathcal{A_{FB}}$ & $< -0.056$ & $< -0.142$ \\
\hline\hline
$B \to K^*_2 \tau^-\mu^+ $ & $ \mathcal{B}$ & $ <9.131 \times 10^{-6}$  & $ <1.416 \times 10^{-6}$ \\
 &$\mathcal{A_{FB}}$ & $< 0.311$ & $< -0.142$ \\
\hline\hline
\end{tabular}
\caption{Upper limits for the branching ratio, forward-backward asymmetry($\mathcal{A_{FB}}$) and lepton polarization ($\mathcal{F_L}$) of $B \to K^*_2 \tau^{\pm}\mu^{\mp}$ LFV decays considering bounds are at $90\%$ C.L..}
\label{tab:resultsk2star}
\end{center}
\end{table}
 \section{Decay distribution of $\Lambda_b \to\Lambda\ell_1^+\ell_2^-$ \label{sec:lambda}}

To establish the kinematics of the decay process, we assume that the $\Lambda_b$ particle is stationary, while the $\Lambda$ and the dilepton pair move in opposite directions along the positive and negative z-axes, respectively. We denote the momenta of these particles as follows: $p$ for the $\Lambda_b$, $k$ for the $\Lambda$, $q_1$ for $\ell_1$, and $q_2$ for $\ell_2$. Additionally, we define $s_p$ and $s_k$ as the spins of the $\Lambda_b$ and $\Lambda$, respectively, with respect to the z-axis in their respective rest frames. We also introduce two key kinematic variables: $q^\mu = q_1^\mu + q_2^\mu$ represents the four-momentum of the dilepton pair, and $\theta_\ell$ represents the angle at which the lepton $\ell_1$ is oriented relative to the z-axis in the rest frame of the dilepton system. The decay amplitudes can be written as
\begin{equation}\label{eq:Mll}
\mathcal{M}^{\lambda_2,\lambda_1}(s_p,s_k) = - \frac{V_{tb}V_{ts}^\ast}{2v^2}\frac{\alpha_e}{4\pi} \sum_{i=L,R}\bigg[ \sum_{\lambda} \eta_\lambda H^{i,s_p,s_k}_{\rm VA, \lambda} L^{\lambda_2,\lambda_1}_{i,\lambda} + H^{i,s_p,s_k}_{\rm SP} L^{\lambda_2,\lambda_1}_i  \bigg]\, .
\end{equation}
In this context, we have two sets of amplitudes: $H^{i,s_p,s_k}_{\rm VA, \lambda}$ and $H^{i,s_p,s_k}_{\rm SP}$. These amplitudes correspond to the vector and axial-vector (VA) operators as well as the scalar and pseudo-scalar (SP) operators. Additionally, we have the leptonic helicity amplitudes denoted as $L^{\lambda_2,\lambda_1}_{i,\lambda}$ and $L^{\lambda_2,\lambda_1}_i$. Here, the subscript $i$ represents the chiralities of the lepton current, and $\lambda$ can take on values of $t$, $+1$, $-1$, or $0$, which correspond to the helicity states of the virtual gauge boson that decays into the dilepton pair. The subscripts $\lambda_{1,2}$ refer to the helicities of the leptons, and $\eta_t$ is equal to 1, while $\eta_{\pm 1,0}$ is equal to $-1$. For the detailed definitions and expressions of $H^{i,s_p,s_k}_{\rm VA, \lambda}$ and $H^{i,s_p,s_k}_{\rm SP}$ in terms of Wilson coefficients and form factors, you can refer to \cite{Das:2018sms}. In the existing literature, transversity amplitudes, denoted as $A^i_{\perp(\|)_{\rm 1}}$, $A^i_{\perp(\|)_{\rm 0}}$, $A_{\rm S\perp(\|)}$, and $A_{\rm P\perp(\|)}$ are frequently used instead of the hadronic helicity amplitudes. You can find the expressions for these transversity amplitudes in Appendix~[\ref{sec:TAs2}] according to the conventions outlined in \cite{Das:2018iap}. 
The  $L^{\lambda_2,\lambda_1}_{i}$ and  $L^{\lambda_2,\lambda_1}_{i,\lambda}$ amplitudes are defined as
\begin{align}
\begin{split}
& L^{\lambda_2,\lambda_1}_{L(R)} = \langle \bar{\ell}_2(\lambda_2)\ell_1(\lambda_1) | \bar{\ell}_2 (1\mp\gamma_5) \ell_1 | 0\rangle\, , \\
\label{eq:Ldef2}
& L^{\lambda_2,\lambda_1}_{L(R),\lambda} = \bar{\epsilon}^\mu(\lambda) \langle \bar{\ell}_2(\lambda_2) \ell_1(\lambda_1) | \bar{\ell}_2 \gamma_\mu (1\mp\gamma_5) \ell_1 | 0\rangle\, .
\end{split}
\end{align}
Here, $\epsilon^{\mu}$ represents the polarization vector of the virtual gauge boson that undergoes decay into the dilepton pair. The specific computations for the values of $L^{\lambda_2,\lambda_1}_{i,\lambda}$ and $L^{\lambda_2,\lambda_1}_{i}$ are provided in Appendix~[\ref{sec:llRF}]. Utilizing these calculations, we determine the differential branching ratio for the decay process $\Lambda_b\to\Lambda\ell_1\ell_2$ as
\begin{equation}\label{eq:twofold}
\frac{d\mathcal{B}}{dq^2 d\cos\theta_\ell} = \frac{3}{2} \bigg( K_{1ss} (1-\cos^2\theta_\ell) + K_{1cc} \cos^2\theta_\ell + K_{1c} \cos\theta_\ell \bigg)\, .
\end{equation}
Each of the angular coefficients $K_{1ss,1cc,1c}$ can be written in the following way:
\begin{equation}
K_{1ss, 1cc} = K_{1ss, 1cc}^{\rm VA} + K_{1ss, 1cc}^{\rm SP} +K_{1ss, 1cc}^{\rm int}\, .
\end{equation} 
The contributions from VA and SP operators are denoted as $K_{1ss, 1cc,1c}^{\rm VA}$ and $K_{1ss, 1cc,1c}^{\rm SP}$, and their interference terms are included in $K_{1ss, 1cc,1c}^{\rm int}$. In terms of the transversity amplitudes, the expressions for $K_{1ss, 1cc,1c}^{\rm VA}$ and $K_{1ss, 1cc,1c}^{\rm SP}$ are as follows:
\begin{align}
&K_{1ss}^{\rm VA} = \frac{1}{4} \bigg( 2|\ARpa0|^2 + |\ARpa1|^2 + 2|\ARpe0|^2 + |\ARpe1|^2 + \{ R \leftrightarrow L  \} \bigg)-  \frac{m_+^2+m_-^2}{4q^2} \nn\\&~~~~~~~ ~~~ \bigg[ \bigg( |A^R_{\|_0}|^2 + |A^R_{\perp_0}|^2 + \{ R \leftrightarrow L \} \bigg) - \bigg( |A_{\perp t}|^2  + \{ \perp \leftrightarrow \| \}\bigg) \bigg] \, + \frac{m_+^2-m_-^2}{4q^2} \nn\\ &~~~~~~~~~ \bigg[ 2\re\bigg( A^R_{\perp_0}A^{\ast L}_{\perp_0} + A^R_{\perp_1}A^{\ast L}_{\perp_1} + \{ \perp \leftrightarrow \| \}  \bigg) \bigg] - \frac{m_+^2m_-^2}{4q^4} \bigg[ \bigg(|\ARpa1|^2 + |\ARpe1|^2 + \{ R \leftrightarrow L \} \bigg)\nn\\& ~~~~~~~~~+ 2|A_{\|t}|^2 + 2|A_{\perp t}|^2\bigg] \, .\\
&K_{1cc}^{\rm VA} = \frac{1}{2}\bigg( |\ARpa1|^2 + |\ARpe1|^2 + \{R \leftrightarrow L \} \bigg) + \frac{m_+^2+m_-^2}{4q^2}\, \bigg[ \bigg( |A^R_{\|_0}|^2 - |A^R_{\|_1}|^2 + |A^R_{\perp_0}|^2 - \nn\\& ~~~~~~~~~|A^R_{\perp_1}|^2 + \{ R \leftrightarrow L \} \bigg)+ \bigg( |A_{\perp t}|^2 + |A_{\| t}|^2 \bigg) \bigg] + \frac{m_+^2-m_-^2}{4q^2} \bigg[ 2\re\bigg( A^R_{\perp_0}A^{\ast L}_{\perp_0}\, + A^R_{\perp_1}A^{\ast L}_{\perp_1} \nn\\&~~~~~~~~ + \{\perp \leftrightarrow \| \} \bigg) \bigg] - \frac{m_+^2 m_-^2}{2q^4}  \bigg[ \bigg(|\ARpa0|^2 + |\ARpe0|^2 + \{ R \leftrightarrow L \} \bigg) + |A_{\|t}|^2 + |A_{\perp t}|^2\bigg] \, . \\
&K_{1c}^{\rm VA} = -\beta_\ell \beta_\ell^\prime \bigg( A^R_{\perp_1}A^{\ast R}_{\|_1} - \{ R \leftrightarrow L \}  \bigg)\, + \beta_\ell \beta_\ell^\prime\frac{m_+ m_-}{q^2}  \re\bigg( \ALpa0 A_{\| t}^\ast + \ALpe0 A_{\perp t}^\ast \bigg)  \, .\\
&K_{1ss}^{\rm SP} = \frac{1}{4}\bigg( |A_{\rm S\perp}|^2 + |A_{\rm P\perp}|^2 + \{ \perp \leftrightarrow \| \} \bigg) - \frac{m_+^2}{4q^2}\big(|A_{S \|}|^2 + |A_{S \perp}|^2\big) - \frac{m_-^2}{4q^2}\big(|A_{P \|}|^2 + |A_{P \perp}|^2\big) \, \nn .~~~~\\ 
\end{align}
\begin{align}
&K_{1cc}^{\rm SP} = \frac{1}{4}\bigg( |A_{\rm P\perp}|^2 + |A_{\rm S\perp}|^2 + \{\perp \leftrightarrow \| \} \bigg)\,- \frac{m_+^2}{4q^2}\big(|A_{S \|}|^2 + |A_{S \perp}|^2\big) - \frac{m_-^2}{4q^2}\big(|A_{P \|}|^2 + |A_{P \perp}|^2\big)\, .\\
&K_{1c}^{\rm SP} = 0\, ,\\.
&K_{1ss}^{\rm int} = \frac{m_+}{2\sqrt{q^2}} \re\bigg(\Apat\AsPpa + \Apet\AsPpe  \bigg) + \frac{m_-}{2\sqrt{q^2}} \re\bigg( \Apat\AsSpa + \Apet\AsSpe \bigg)- \,\nn\\&~~~~~~~~\frac{m_+^2m_-}{2q^2\sqrt{q^2}} \re\bigg( \Apat\AsSpa +  \Apet\AsSpe \bigg) - \frac{m_+m_-^2}{2q^2\sqrt{q^2}}\re\bigg( \Apat\AsPpa +  \Apet\AsPpe \bigg)\, .\\
&K_{1cc}^{\rm int} = \frac{m_+}{2\sqrt{q^2}}\re\bigg(\Apat\AsPpa + \Apet\AsPpe \bigg) + \frac{m_-}{2\sqrt{q^2}}\re\bigg(\Apat\AsSpa + \Apet\AsSpe \bigg) - \,\nn\\&~~~~~~~~~\frac{m_+^2m_-}{2q^2\sqrt{q^2}}\re\bigg( \Apat\AsSpa + \Apet\AsSpe \bigg)\,- \frac{m_+ m_-^2}{2q^2\sqrt{q^2}}\re\bigg( \Apat\AsPpa + \Apet\AsPpe \bigg)\, .\\
&K_{1c}^{\rm int} = \frac{\beta_\ell\beta_\ell^\prime}{2\sqrt{q^2}} \re\bigg( \ASpa\AsLpa{0} + \ASpe\AsLpe{0} + \ASpa\AsRpa{0} + \ASpe\AsRpe{0} \bigg) \nn\\ &~~~~~~~~+ \frac{\beta_\ell\beta_\ell^\prime}{2\sqrt{q^2}} \re\bigg( \APpa\AsLpa{0} + \APpe\AsLpe{0} - \APpa\AsRpa{0} - \APpe\AsRpe{0} \bigg) .
\end{align} 
We've introduced the notations $m_\pm = m_1 \pm m_2$, where $m_1$ and $m_2$ represent the masses of $\ell_1$ and $\ell_2$ respectively. The coefficients $\beta_{\ell}^{(\prime)}$ are specified in Appendix [\ref{sec:TAs2}]. Based on the differential decay distribution, we establish two observables, as outlined in \cite{Das:2018iap}: the differential branching ratio is
\begin{equation} \label{eq:diffBr}
	\frac{d\mathcal{B}}{dq^2} = 2 K_{1ss} + K_{1cc}\, ,
\end{equation}
and the forward-backward asymmetry 
\begin{equation}\label{eq:AlFB}
A^{\ell}_{\rm FB} = \frac{3}{2} \frac{K_{1c}}{K_{1ss} + K_{1cc}}\, .
\end{equation}
The available phase space in the Dilepton invariant mass squared $q^2$ is
\begin{equation}
	(m_1+m_2)^2 \le q^2 \le (\mLb - \mL)^2\, .
\end{equation}
\subsection{Numerical Analysis for $\Lambda_b \to\Lambda\ell_1\ell_2$ }

Given that the mesonic decays $B^+ \to K^+ \ell_1 \ell_2$ and $\bar{B}_s \to \ell_1 \ell_2$ arise from the same underlying quark-level transition as the baryonic mode $\Lambda_b \to \Lambda \ell_1 \ell_2$, we utilize the constrained NP parameter space, obtained from these mesonic channels and illustrated in Figure~\ref{fig::ModelIndep-constrainstau} to study the corresponding baryonic decays. Specifically, we investigate the $q^2$ dependence of the differential branching ratio and the forward-backward asymmetry for the LFV processes $\Lambda_b \to \Lambda \tau^\pm \mu^\mp$.

\begin{figure}[htb]
\begin{tabular}{cc}
	\centering
	\includegraphics[width=0.48\textwidth]{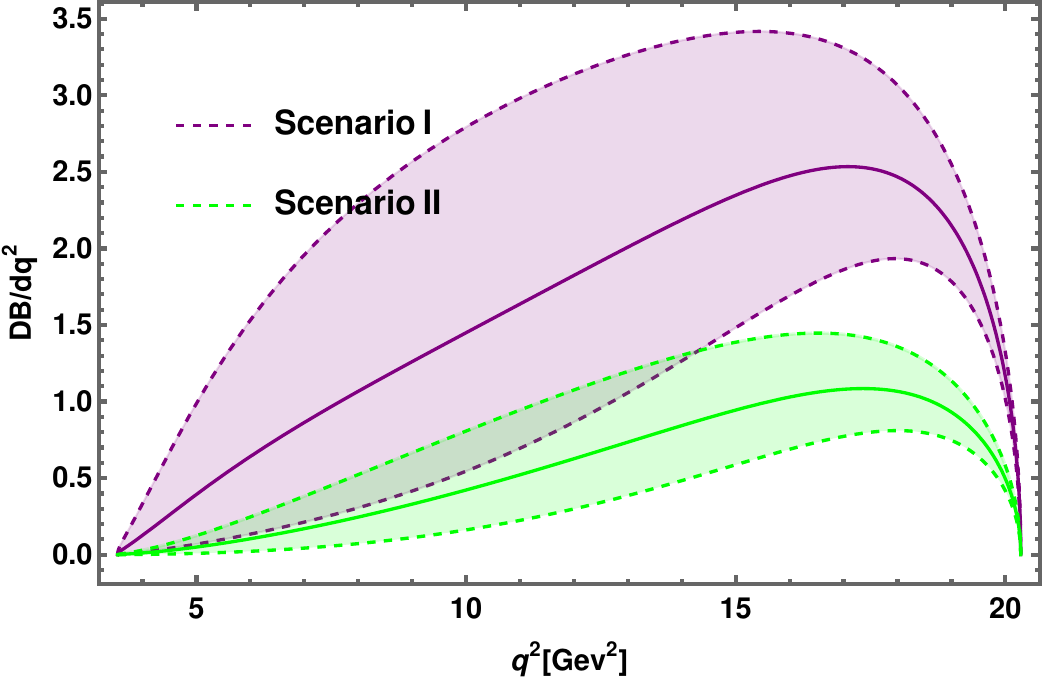} &
  	\includegraphics[width=0.48\textwidth]{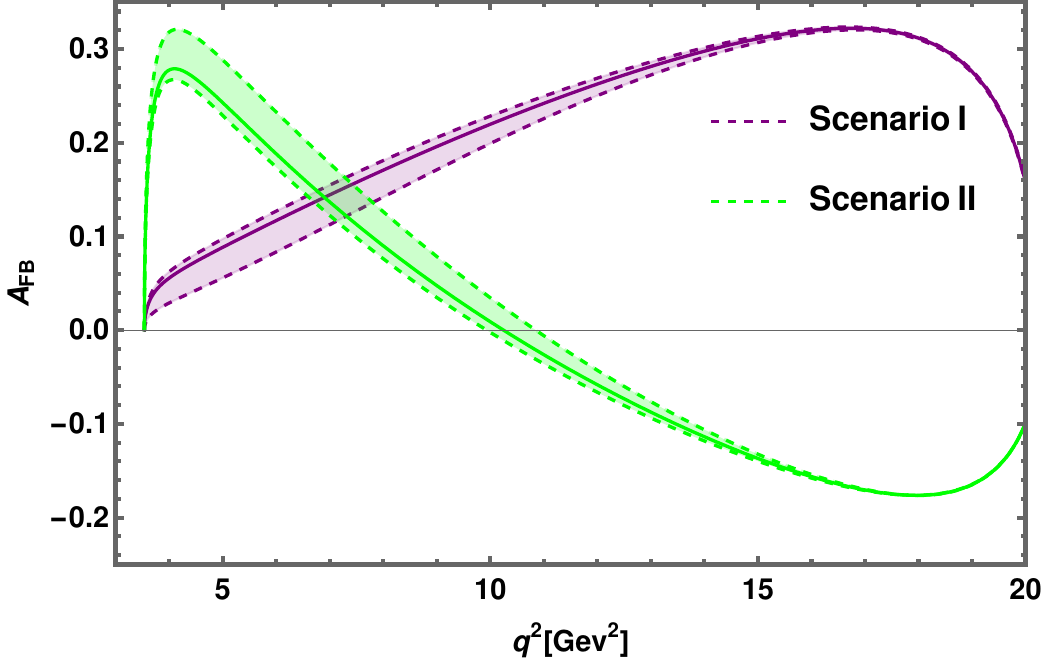}\\
  	\end{tabular}
    \caption{ Branching ratio (in units of $10^{-6}$)  (left) and forward backward asymmetry (right) for $\Lambda_b \to\Lambda \tau^+\mu^-$.}\label{fig::lambdataupmum}
    \end{figure}

\begin{figure}[htb]
\begin{tabular}{cc}
	\centering
	\includegraphics[width=0.48\textwidth]{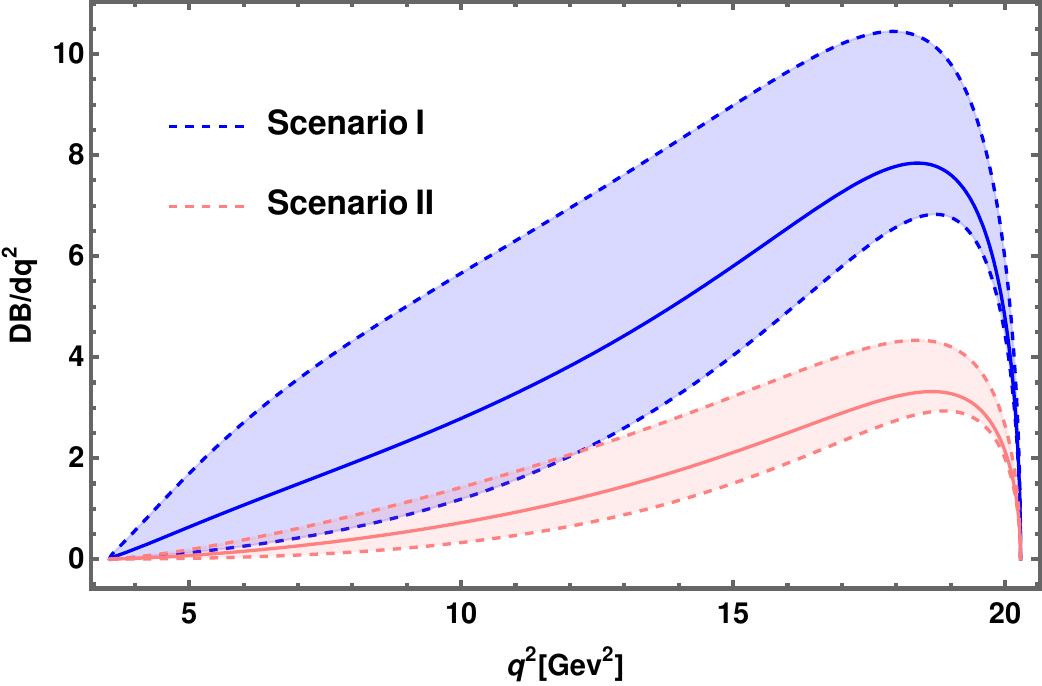} &
  	\includegraphics[width=0.48\textwidth]{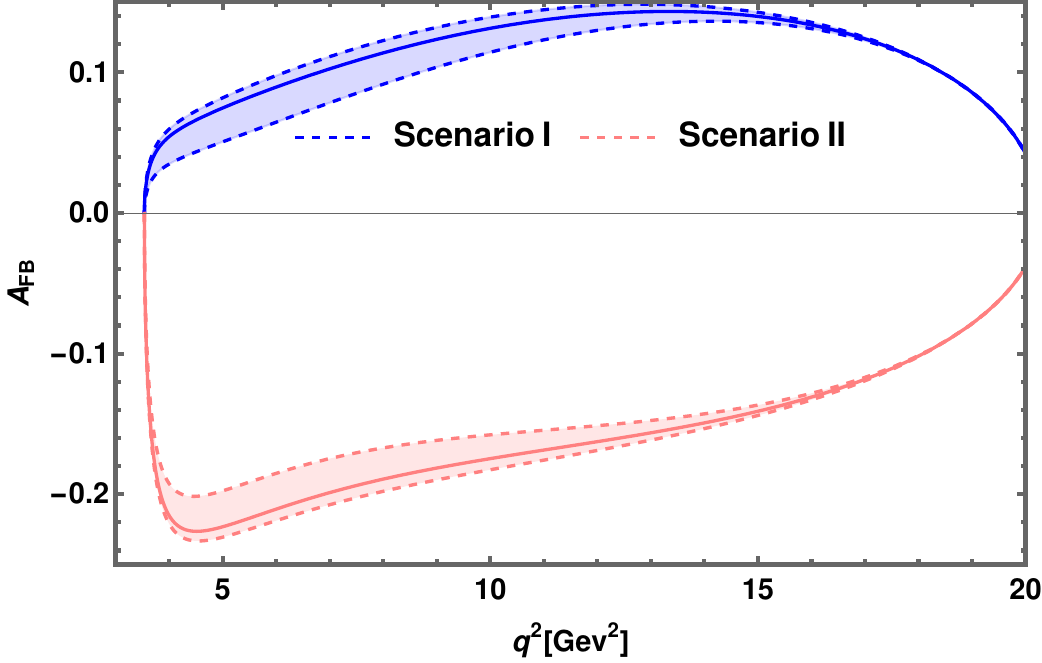}\\
  	\end{tabular}
    \caption{ Branching ratio (in units of $10^{-6}$) (left) and forward backward asymmetry (right) for $\Lambda_b \to\Lambda \tau^-\mu^+$.}\label{fig::lambdataummup}
    \end{figure}

Figure~\ref{fig::lambdataupmum} displays the $q^2$ distributions for the $\Lambda_b \to \Lambda \tau^+ \mu^-$ decay. The left panel presents the differential branching ratio $\mathrm{d}\mathcal{B}/\mathrm{d}q^2$, while the right shows the forward-backward asymmetry $\mathcal{A}_{\text{FB}}(q^2)$, comparing Scenario~I (cyan blue) and Scenario~II (magenta), each with their associated $1\sigma$ uncertainty bands. Similarly, Figure~\ref{fig::lambdataummup} presents the corresponding observables for the $\Lambda_b \to \Lambda \tau^- \mu^+$ decay, where Scenario~I is shown in purple and Scenario~II in orange.

The branching ratios of the LFV decays $\Lambda_b \to \Lambda \tau^\pm \mu^\mp$ demonstrate substantial deviations , reaching values up to $\mathcal{O}(10^{-5})$, thereby serving as clear signatures of LFV. In the $\tau^+ \mu^-$ mode, Scenario~I dominates at low $q^2$, whereas Scenario~II becomes prominent at intermediate $q^2$, suggesting contributions from heavier particles. A gradual increase in the branching ratio is observed with increasing $q^2$, peaking around $q^2 \sim 3.5$ in Scenario~II. The divergence between the two scenarios becomes more noticeable beyond $q^2 \approx 10~\text{GeV}^2$, indicating a strong dependence on the NP operator structure.

The  $\mathcal{A}_{\text{FB}}(q^2)$ also reveals notable differences. For both decay modes, Scenario~I exhibits a consistently positive asymmetry that peaks around $q^2 \sim 10$--$12~\text{GeV}^2$ before tapering off at the kinematic extremes. In contrast, Scenario~II shows a zero-crossing behavior in the $\tau^+ \mu^-$ decay and remains entirely negative in the $\tau^- \mu^+$ mode. These distinctions emphasize the asymmetry's sensitivity to the nature and chirality of the underlying NP interactions, with intermediate $q^2$ regions being particularly informative.

The quantitative results for the integrated branching ratios and forward-backward asymmetries are summarized in Table~\ref{tab:resultslambda}. These values incorporate the upper bound constraints on LFV branching ratios at the $90\%$ confidence level and offer a detailed benchmark for future experimental investigations into baryonic LFV processes.

  \begin{table}[h!]
\begin{center}
\renewcommand{\arraystretch}{1.2} 
\begin{tabular}{||c| c| c| c||}
\hline\hline
\textbf{Decay modes}& \textbf{Observables} & \textbf{Scenario I}  & \textbf{Scenario II} \\
 \hline
$\Lambda_b\to\Lambda \tau^+\mu^-$ & $ \mathcal{B} $ & $ <2.63 \times 10^{-5}$  & $ <1.673\times 10^{-5}$ \\
  &$\mathcal{A_{FB}}$ & $< 0.223$ & $< -0.012$ \\
\hline\hline
$\Lambda_b\to\Lambda  \tau^-\mu^+ $ & $ \mathcal{B}$ & $ <6.583\times 10^{-5}$  & $ <9.631 \times 10^{-5}$ \\
   & $\mathcal{A_{FB}}$ & $< 0.111$ & $< -0.160 $\\
\hline\hline
\end{tabular}
\caption{Upper limits for the branching ratio, forward-backward asymmetry($\mathcal{A_{FB}}$) and lepton polarization ($\mathcal{F_L}$) of $\Lambda_b\to\Lambda \tau^{\pm}\mu^{\mp}$ LFV decays considering bounds are at $90\%$ C.L..}
\label{tab:resultslambda}
\end{center}
\end{table}
 \section{Summary \label{sec:summary}}  
In this work, we have investigated a range of LFV decays: $B \rightarrow K^* \ell_1 \ell_2$, $B \rightarrow \phi \ell_1 \ell_2$, $B \rightarrow K^*_2 \ell_1 \ell_2$, and $\Lambda_b \rightarrow \Lambda \ell_1 \ell_2$. Our study considers contributions from vector, axial-vector, scalar, and pseudo-scalar NP operators and examines two distinct scenarios under these assumptions. We focus on the double-differential decay distribution in terms of the dilepton invariant mass squared ($q^2$) and the lepton angle ($\theta_\ell$), from which we extract observables such as the differential branching ratio, forward-backward asymmetry, and lepton polarization. The analysis is performed in a model-independent framework, allowing us to set upper bounds on the branching fractions and $\mathcal{A}_{FB}$ at the 90\% confidence level, with potential sensitivity for experimental verification at LHCb. Our findings indicate that the branching ratios for the decays $B \to (K^*, \phi)\, \tau^{\pm} \mu^\mp$ are of the order $\mathcal{O}(10^{-5})$, with negligible difference between charge-conjugate channels. For the $B \to K^*_2\, \tau^\pm \mu^\mp$ modes, the branching ratio is smaller, at $\mathcal{O}(10^{-6})$. We also identify the zero-crossing points of $\mathcal{A}_{FB}$ in the $B \to (K^*, \phi, K^*_2)\, \mu^+ \tau^-$ modes. In the baryonic sector, the branching ratios for $\Lambda_b \rightarrow \Lambda \tau^+ \mu^-$ and $\Lambda_b \rightarrow \Lambda \tau^- \mu^+$ are estimated to be $2.63 \times 10^{-5}$ and $6.58 \times 10^{-5}$, respectively. Notably, the forward-backward asymmetry exhibits a zero-crossing only in the $\tau^+ \mu^-$ mode, suggesting potential CP-violating or chiral NP effects. This charge asymmetry implies non-trivial structures in the LFV couplings, offering a promising window into distinguishing NP interactions by their lepton flavor and charge dependence.
In summary, our analysis underscores the value of LFV observables as sensitive probes of new physics and highlights several decay channels as key targets for ongoing and future experimental investigations.

\section*{Acknowledgements}

AB expresses sincere gratitude to Prof. Sanjay Kumar Swain for his valuable guidance and moral support, as well as for the financial assistance provided by the NISER Planned Project (Project No. RIN-4001). DP acknowledges the financial support from the Prime Minister’s Research Fellowship, Government of India. RM gratefully acknowledges the University of Hyderabad for supporting this work through the IoE project grant (Grant No. RC1-20-012).

\appendix
\section{ Angular conventions and Kinematics}
\begin{figure}[htb]
\begin{center}
\includegraphics[scale=0.58]{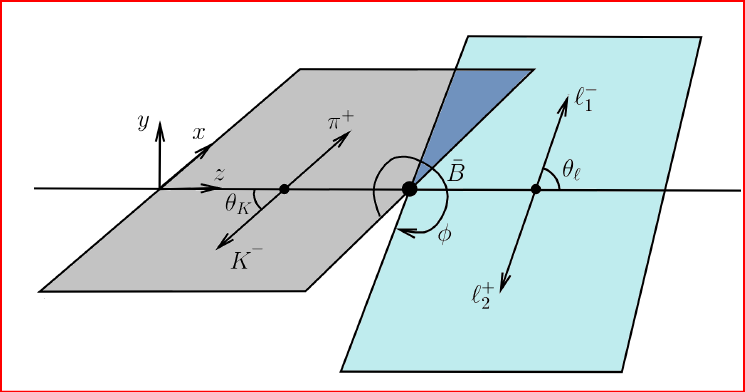}
\end{center}
\caption{\label{fig:angles} \small Angular conventions for the decay $\bar{B}\to \bar{K}^\ast \ell_1^- \ell_2^+$.}
\end{figure}

Our angular conventions for the decay $\bar{B}\to\bar{K}^\ast(\to K^- \pi^+)\ell_1^-\ell_2^+$ are summarized in Fig.~\ref{fig:angles}. 
In the $B$ rest frame, the leptonic and hadronic four-vectors are defined by $q^\mu=(q_0,0,0,q_z)$ and $k^\mu=(k_0,0,0,-q_z)$, where 
\begin{equation}
q_0=\frac{m_B^2+q^2-m_{K^\ast}^2}{2 m_B},\qquad k_0=\frac{m_B^2+m_{K^\ast}^2-q^2}{2 m_B},\quad\text{and}\quad q_z=\frac{\lambda^{1/2}(m_B,m_{K^\ast},\sqrt{q^2})}{2 m_B},
\end{equation}
In the dilepton rest frame, the leptonic four-vectors read
\begin{align}
p_{1}^\mu &= (E_\alpha, |p_\ell| \sin\theta_\ell \cos\phi,-|p_\ell| \sin \theta_\ell \sin\phi,|p_\ell|\cos\theta_\ell),\\
p_{2}^\mu &= (E_\beta, -|p_\ell| \sin\theta_\ell \cos\phi,|p_\ell| \sin \theta_\ell \sin\phi,-|p_\ell|\cos\theta_\ell),
\end{align}
where 
\begin{equation}
E_1=\frac{q^2+m_1^2-m_2^2}{2 \sqrt{q^2}},\qquad E_2=\frac{q^2-m_1^2+m_2^2}{2 \sqrt{q^2}},\quad\text{and}\quad |p_\ell|=\frac{\lambda^{1/2}(q^2,m_1^2,m_2^2)}{2 m_B}.
\end{equation}
In the same way, one can write in the $K^\ast$ rest frame
\begin{align}
p_K^\mu &= (E_K, -|p_K|\sin \theta_K,0,-|p_K|\cos\theta_K),\\
p_\pi^\mu &= (E_\pi,+|p_K|\sin \theta_K,0,+|p_K|\cos\theta_K),
\end{align}
where $E_K$, $E_\pi$ and $|p_K|$ are given by the similar expressions.

\section{Polarization of $K^*_2$}
\label{appenb}
The polarization $\epsilon^{\mu\nu}(n)$ of tensor meson $K_2^\ast$, which has four momentum $(k_0, 0, 0, \vec{k})$, can be written in terms of the spin-1 polarization vectors~\cite{Berger:2000wt}
\begin{eqnarray}
 \epsilon_{\mu\nu}(\pm 2) &=& \epsilon_\mu(\pm 1)\epsilon_\nu(\pm 1),\nn\\
 \epsilon_{\mu\nu}(\pm 1) &=& \frac{1}{\sqrt{2}}\left[\epsilon_\nu(\pm)\epsilon_\nu(0) + \epsilon_\nu(\pm)\epsilon_\mu(0) \right],\nn\\
 \epsilon_{\mu\nu}(0) &=& \frac{1}{\sqrt{6}}\left[\epsilon_\mu(+) \epsilon_\nu(-) + \epsilon_\nu(+) \epsilon_\mu(-) \right]+ \sqrt{\frac{2}{3}}\epsilon_\mu(0)\epsilon_\nu(0) ,
 \end{eqnarray} 
where the spin-1 polarization vectors are defined as
\begin{equation}
\epsilon_\mu(0) = \frac{1}{m_{K_2^\ast}}\left(\vec{k}_z,0,0,k_0\right)\, ,\quad \epsilon_\mu(\pm) = \frac{1}{\sqrt{2}}\left(0,1,\pm i, 0\right)\
\end{equation}

We study $B\to K^*_2\ell_1\ell_2$ decay where we have two leptons in the final state so in this case, the $n=\pm 2$ helicity states of the $K_2^\ast$
is not realized. Therefore a new polarization vector is introduced~\cite{Wang:2010tz}
\begin{equation}
\epsilon_{T\mu}(h) = \frac{\epsilon_{\mu\nu}p^\nu}{m_B}\, 
\end{equation}
 The explicit expressions of polarization vectors are
\begin{eqnarray}
\epsilon_{T\mu}(\pm 1) &=& \frac{1}{m_B}\frac{1}{\sqrt{2}}\epsilon(0).p  \epsilon_\mu(\pm) = \frac{\sqrt{\lambda}}{\sqrt{8}m_B m_{K^*_2}} \epsilon_\mu(\pm), \\
\epsilon_{T\mu}(0) &=& \frac{1}{m_B}\sqrt{\frac{2}{3}}\epsilon(0).p \epsilon_\mu(0) = \frac{\sqrt{\lambda}}{\sqrt{6}m_B m_{K^*_2}} \epsilon_\mu(0),
\end{eqnarray}
where $\lambda(m^2_B,m^2_{K^*_2},q^2) = m^4_B + m^4_{K^*_2} + q^4 -2(m^2_B m^2_{K^*_2}+m^2_Bq^2+m^2_{K^*_2}q^2)$ is the usual Kallen function.
On the other hand, the virtual gauge boson can have three types of polarization states, longitudinal, transverse and time-like, which have following components
\begin{equation}
\epsilon^\mu_V(0) = \frac{1}{\sqrt{q^2}}(-|\vec{q_z}|,0,0,-q_0)\, ,\quad \epsilon^\mu_V(\pm) = \frac{1}{\sqrt{2}}(0,1,\pm i, 0)\ ,\quad
\epsilon^\mu_V(t) = \frac{1}{\sqrt{q^2}}(q_0,0,0,q_z)\ 
\end{equation}
where $q^\mu=(q_0,0,0,q_z)$ is four momentum of gauge boson.
\section{Transversity amplitudes \label{sec:TAs2}}
Corresponding to the effective Hamiltonian \eqref{eq:Heff1} the expressions of the transversity amplitudes read \cite{Das:2018iap}
\begin{eqnarray}
A^{L,(R)}_{\perp_1} &=& -\sqrt{2}N \bigg( f^V_\perp \sqrt{2s_-} \mC^{L,(R)}_{\rm VA+} \bigg)\, ,\\
A^{L,(R)}_{\|_1} &=& \sqrt{2}N \bigg( f^A_\perp \sqrt{2s_+} \mC^{L,(R)}_{\rm VA-}  \bigg)\, ,\\
A^{L,(R)}_{\perp_0} &=& \sqrt{2}N \bigg( f^V_0 (\mLb + \mL) \sqrt{\frac{s_-}{q^2}} \mC^{L,(R)}_{\rm VA+}  \bigg)\, ,\\
A^{L,(R)}_{\|_0} &=& -\sqrt{2}N \bigg( f^A_0 (\mLb - \mL) \sqrt{\frac{s_+}{q^2}} \mC^{L,(R)}_{\rm VA-}  \bigg)\, ,\\
A_{\perp t} &=& -2\sqrt{2}N f^V_t (\mLb - \mL) \sqrt{\frac{s_+}{q^2}} (\mC_{A} + \mC_A^\prime)\, ,\\
A_{\| t} &=& 2\sqrt{2}N f^A_t (\mLb + \mL) \sqrt{\frac{s_-}{q^2}} (\mC_{A} - \mC_A^\prime) \, .
\end{eqnarray}
Here the normalization constant $N(q^2)$ is given by 
\begin{align}
&N(q^2) =  \frac{V_{tb}V_{ts}^\ast \alpha_e}{\sqrt{2}v^2} \sqrt{ \tau_{\Lambda_b} \frac{q^2 \sqrt{\lambda(\mmLb,\mmL,q^2)} }{2^{15} m^3_{\Lambda_b} \pi^5}	\beta_\ell \beta_\ell^\prime}\, ,\nn\\
&\beta_\ell = \sqrt{1 - \frac{(m_1+m_2)^2}{q^2}}\,,\quad \beta_\ell^\prime = \sqrt{1 - \frac{(m_1-m_2)^2}{q^2}}\, ,
\end{align}
where $\lambda(a,b,c)=a^2 + b^2 + c^2 - 2(ab + bc + ca)$ and the Wilson coefficients are
\begin{align}
& \mC_{\rm VA,+}^{L(R)} = (\mC_V\mp \mC_A)+(\mC_V^\prime \mp \mC_A^\prime)\, ,\\
& \mC_{\rm VA,-}^{L(R)} = (\mC_V\mp \mC_A)-(\mC_V^\prime \mp \mC_A^\prime)\, .
\end{align}
The transversity amplitudes corresponding to the SP operators are \cite{Das:2018iap}
\begin{eqnarray}
A_{\rm S\perp} &=& 2\sqrt{2}N f^V_t \frac{\mLb - \mL}{m_b} \sqrt{s_+} (\mC_S + \mC_S^\prime)\, ,\\
A_{\rm S\|} &=& -2\sqrt{2}N f^A_t \frac{\mLb + \mL}{m_b} \sqrt{s_-} (\mC_S - \mC_S^\prime)\, ,\\
A_{\rm P\perp} &=& -2\sqrt{2}N f^V_t \frac{\mLb - \mL}{m_b} \sqrt{s_+} (\mC_P + \mC_P^\prime)\, ,\\
A_{\rm P\|} &=& 2\sqrt{2}N f^A_t \frac{\mLb + \mL}{m_b} \sqrt{s_-} (\mC_P - \mC_P^\prime)\, .
\end{eqnarray}
\section{Spinors in dilepton rest frame \label{sec:llRF}}
We assume that the lepton $\ell_2^-$ is negatively charged and has four-momentum is $q_2^\mu=(E_1,\vec{q})$, while $\ell_1^+$ is positively charged and has four-momentum $q_1^\mu=(E_1,-\vec{q})$ 
\begin{align}
q_1^\mu \Big|_{2\ell}&  = (E_2, -|q_{2\ell}|\sin\theta_\ell, 0, -|q_{2\ell}|\cos\theta_\ell)\, ,\\
q_2^\mu \Big|_{2\ell}& = (E_1, |q_{2\ell}|\sin\theta_\ell, 0, |q_{2\ell}|\cos\theta_\ell)\, ,
\end{align}
with 
\begin{eqnarray}
|q_{2\ell}| &=& \frac{\lambda^{1/2}(q^2,m_1^2,m_2^2)}{2\sqrt{q^2}}\, ,\quad\quad E_1 = \frac{q^2+m_1^2-m_2^2}{2\sqrt{q^2}}\,,\nn\\  E_2 &=& \frac{q^2+m_2^2-m_1^2}{2\sqrt{q^2}}\, .
\end{eqnarray}
The explicit expressions of the lepton helicity amplitudes require us to calculate 
\begin{equation}
\bar{u}_{\ell_2} (1\mp\gamma_5) v_{\ell_1}\, ,\quad \bar{\epsilon}^\mu(\lambda) \bar{u}_{\ell_2} \gamma_\mu (1\mp\gamma_5) v_{\ell_1}\, .
\end{equation}
With these choices of lepton spinors we get the following expressions of the lepton helicity amplitudes $L^{\lambda_2,\lambda_1}_{L(R)}$ and $ L^{\lambda_2,\lambda_1}_{L(R),\lambda}$
\begin{align}
&L_L^{\plpl} = \sqrt{q^2}(\beta_\ell^\prime+\beta_\ell)\, ,\quad L_L^{\plmi} = 0\, ,\quad L_L^{\mipl} = 0\,  ,\quad  L_L^{\mimi} = \sqrt{q^2}(\beta_\ell^\prime-\beta_\ell)\, ,\\
& L_R^{\plpl} = -\sqrt{q^2}(\beta_\ell^\prime - \beta_\ell)\, ,\quad L_R^{\plmi} = 0\, ,\quad L_R^{\mipl} = 0\, ,\quad L_R^{\mimi} = -\sqrt{q^2}(\beta_\ell^\prime + \beta_\ell) ,\\
& L_{L,+1}^{\plpl} = \frac{1}{\sqrt{2}}\big[m_1(\beta_\ell^\prime +\beta_\ell) + m_2(\beta_\ell^\prime - \beta_\ell) \big] ,\quad  L_{L,+1}^{\plmi} = -\sqrt{\frac{q^2}{2}} (\blp-\bl) (1-\cl)\, ,\nn\\
& L_{L,+1}^{\mimi} = -\frac{1}{\sqrt{2}}\big[m_1(\blp - \bl) + m_2(\blp + \bl) \big] ,\quad L_{L,+1}^{\mipl} = \sqrt{\frac{q^2}{2}}(\blp+\bl)(1+\cl)   ,\\
& L_{R,+1}^{\plpl} = \frac{1}{\sqrt{2}}\big[ m_1(\blp-\bl) + m_2(\blp+\bl) \big]  ,\quad L_{R,+1}^{\plmi} = -\sqrt{\frac{q^2}{2}} (\blp+\bl)(1-\cl)\, ,\nn\\
 & L_{R,+1}^{\mimi} = -\frac{1}{\sqrt{2}} \big[ m_1(\blp+\bl) + m_2(\blp-\bl) \big] ,\quad 
L_{R,+1}^{\mipl} = \sqrt{\frac{q^2}{2}} (\blp-\bl) (1+\cl)\,  \\
& L_{L,-1}^{\plpl} = -\frac{1}{\sqrt{2}} \big[m_1(\blp+\bl)+m_2(\blp-\bl) \big]\,  ,\quad  L_{L,-1}^{\plmi} = -\sqrt{\frac{q^2}{2}} (\blp-\bl) (1+\cl)\, ,\nn\\
& L_{L,-1}^{\mimi} = \frac{1}{\sqrt{2}} \big[m_1(\blp-\bl) + m_2(\blp+\bl) \big] ,\quad  L_{L,-1}^{\mipl} = \sqrt{\frac{q^2}{2}} (\blp+\bl) (1-\cl) ,
\end{align}
\begin{align}
& L_{R,-1}^{\plpl} = -\frac{1}{\sqrt{2}} \big[m_1(\blp-\bl) + m_2(\blp+\bl) \big]  ,\quad  L_{R,-1}^{\plmi} = \sqrt{\frac{q^2}{2}}(\blp+\bl)(1+\cl)\,  ,\nn\\
& L_{R,-1}^{\mimi} = \frac{1}{\sqrt{2}}\big[m_1(\blp+\bl) + m_2(\blp-\bl) \big]  ,\quad L_{R,-1}^{\mipl} = \sqrt{\frac{q^2}{2}}(\blp-\bl)(1-\cl)\, ,\\
& L_{L,0}^{\plpl} = -\big[m_1(\blp+\bl)-m_2(\blp-\bl) \big]\cl\, ,\quad L_{L,0}^{\plmi} = \sqrt{q^2}(\blp-\bl)\cl\,,\nn\\
& L_{L,0}^{\mimi} = \big[m_1(\blp-\bl) + m_2(\blp+\bl) \big]\cl\, ,\quad L_{L,0}^{\mipl}=\sqrt{q^2}(\blp+\bl)\cl\, ,\\
& L_{R,0}^{\plpl} = -\big[m_1(\blp-\bl) + m_2(\blp+\bl) \big]\cl\,  ,\quad L_{R,0}^{\plmi} = \sqrt{q^2}(\blp+\bl)\cl\, ,\nn\\
& L_{R,0}^{\mimi} = \big[m_1(\blp+\bl) + m_2(\blp-\bl) \big]\cl\, ,\quad L_{R,0}^{\mipl} =\sqrt{q^2}(\blp-\bl) \\
& L_{L,0}^{\plpl} = \big[m_1(\blp+\bl) + m_2(\blp-\bl) \big]\, ,\nn\\ &L_{L,0}^{\plmi} = L_{L,0}^{\mipl} = 0\, ,\quad L_{L,0}^{\mimi} = \big[m_1(\blp-\bl) + m_2(\blp+\bl) \big]\, ,\\
& L_{R,0}^{\plpl} = -\big[m_1(\blp-\bl) + m_2(\blp+\bl) \big]\, ,\nn\\&L_{R,0}^{\plmi} = L_{R,0}^{\mipl} = 0\, ,\quad  L_{R,0}^{\mimi} = -\big[m_1(\blp+\bl) + m_2(\blp-\bl) \big]\, .
\end{align}

\vspace{3cm}
\bibliographystyle{ieeetr}
\bibliography{ref}

\end{document}